\documentclass[letterpaper,11pt]{article}
\pdfoutput=1
\usepackage{jheppub}
\usepackage{booktabs} 
\usepackage{graphicx} 
\usepackage{siunitx}
\usepackage{comment}
\usepackage{setspace}
\usepackage{amssymb}
\usepackage{amsmath}
\usepackage{comment} 
\usepackage{amstext}
\usepackage{amsfonts}
\usepackage{bbold}
\usepackage{slashed}
\usepackage{tikz-feynman}
\usepackage{xcolor}
\usepackage{subcaption}
\usepackage{physics}
\usepackage{slashed}
\newcommand{\beq}{\begin{equation}}
\newcommand{\eeq}{\end{equation}}
\def\l{\left}
\def\r{\right}

\def\bea{\begin{eqnarray}}
\def\eea{\end{eqnarray}}

\def\nn{\nonumber}
\def\nbar{\bar n}
\def\bn{\bar n}

\def\be{\begin{eqnarray}}
\def\ee{\end{eqnarray}}

\definecolor{SGreen}{rgb}{0.0547,0.613,0.328}
\definecolor{NodeBlue}{rgb}{0.0547,0.148,0.578}

\newcommand{\dbar}{d\hspace*{-0.08em}\bar{}\hspace*{0.1em}}
\begin{document}

\title{Extracting the Asymptotic Behavior of S-matrix Elements from their Phases} 

\author{Ira Z. Rothstein and}
\author{Michael Saavedra}
\affiliation{Department of Physics, Carnegie Mellon University, Pittsburgh, PA 15213, USA}

\abstract{

The asymptotic kinematic limits of S-matrices are dominated by large logarithms which, roughly speaking, fall into two categories: those which are controlled by a renormalization group (RG) scale, which we may think of as logs involving ratios of
invariant mass scales, and those which are functions of ratios of rapidities, so called ``rapidity logs". It has been pointed out by Caron-Huot and Wilhlem \cite{Caron-Huot2016RenormalizationS-matrix} that RG anomalous dimension can be extracted from the phase of the S-matrix, which can greatly simplify calculations
via unitarity methods. 
In this paper we generalize the results of \cite{Caron-Huot2016RenormalizationS-matrix} to show that the phase can be used to reconstruct rapidity anomalous dimensions, by performing a special type of  complex boost. The methodology introduced allows one  to calculate without the need for a rapidity regulator which can  lead to significant simplifications. We demonstrate the use of this method to derive the rapidity anomalous dimensions in the Sudakov form factor and the two parton soft function at two loops order.}

\maketitle

\section{Introduction}
Canonical perturbation theory inadequately describes field theories when dimensionful parameters
form large hierarchies that lead to numerically large logarithms. Typically, perturbation theory can be
salvaged by re-summing these logs using the renormalization group (RG) which takes advantage of
the invariance of physical quantities under the change in subtraction scales.
There exists another class of logs, which may be with associated with IR divergences,
that are not immediately summable by naive RG methods. For instance, often we run into logs whose argument
involves a mass(es), e.g. $\ln(p^2/M^2)$. These logs {\it may} be summable using the RG if the
masses are acting as ``intermediate" scales, in the sense that there is some physical IR scale
below the scale of the mass. In such cases one can work within an EFT and integrate out
the mass. Below the mass scale the IR divergences get converted to a UV divergence and the
logs become amenable to canonical RG methods. 
Effectively what this process amounts to is encapsulated in the following relation
\beq
\label{RG}
\ln(p^2/M^2)= \ln(p^2/\mu^2)-\ln(M^2/\mu^2)
\eeq
where the $RG$ scale $\mu$, is taking on double duty as both the invariant mass factorization and RG scales.
Any log which is summable by RG can be considered an invariant mass log, as the RG flow
corresponds to a Wilsonian flow in the invariant mass \footnote{It is important to emphasize here that the logs of interest in this paper are only large when working in Minkowskian signature.}. Notice that the RG is a manifestation of
factorization. That is, the RG scales distinguishes between low and high virtuality modes such that
an amplitude can be written as a product 
\beq{ \cal{M}}= H(Q,\mu)S(p_i,\mu)
\eeq
where $Q$ is the short distance (hard) scale and $p_i$ are the small momenta.  The log of the mass in eq. (\ref{RG}) goes into the hard function while the other log is part of the soft. This is nothing more 
than an operator product expansion. The key point we are reviewing here is that the logs are summable
because amplitudes factorize in invariant mass.

There exists another category of logs which are not summable using traditional RG methods. In particular, one may have logs associated
with a large ratio in kinematic variables, such as a Regge log of the form $\ln(s/t)$. 
Naively this may seem like a large ratio of virtualities. But this would be an erroneous conclusion.
Such logs arise from forward scattering amplitudes where the large momentum {\it never runs through the loops}.
While the diagram is not (directly) sensitive to the large invariant mass, $s$, it is
sensitive to the large light cone fractions carried by the incoming particles.
The existence of this log is more readily understood in terms of a ratio of
rapidities by writing, in analogy to (\ref{RG}),
\beq
\label{RRG}
\ln(s/t)= \ln(s/\nu^2)-\ln(t/\nu^2)=\ln(p_+/\nu)+\ln(p_-/\nu)-\ln(t/\nu^2),
\eeq
where $s=p_+p_-$ and $\nu$ is a ``rapidity factorization scale" that is analogous to
the RG scale $\mu$. Once the logs are split in this way one can sum them using
the ``Rapidity Renormalization Group" (RRG) \cite{Chiu:2011qc,Chiu2012ATheory}, in a similar way to the canonical RG.
However, to perform the resummation we need to show that the amplitude factorizes in rapidity
space in analogy with the factorization in invariant mass. 
Once this is done one can resum the rapidity logs by calculating the Rapidity Anomalous Dimension (RAD), which is the analog of the usual
RG anomalous dimension.
A crucial distinction between the RRG and RG, is that the coupling does not depend upon the rapidity
factorization scale $\nu$. Thus, the full amplitude is independent of $\nu$ exactly, at each order in the coupling


The RG anomalous dimension is determined by calculating the UV poles arising from 
operator insertions and following the Feynman rules from a given Lagrangian. 
Reference \cite{Caron-Huot2016RenormalizationS-matrix} noted that the RG anomalous dimensions 
could be calculated using unitarity/on-shell methods that have facilitated modern higher order radiative corrections. 
They  showed that the RG anomalous dimensions 
are intimately related to the phase of the
S-matrix, essentially as a consequence of the fact that the imaginary part of the amplitude is the discontinuity in logarithms. 
A general log in an amplitude relates the large log to the phase via
\beq
A\ln(- q^2 /m^2-i\epsilon)= A\ln( q^2/m^2)-i A \pi.
\eeq 
This fact is, not only formally interesting, but also leads to technical
simplifications as unitarity methods can be used to effectively ``gain a loop" in the sense that one can calculate using cut diagrams \cite{Caron-Huot2016RenormalizationS-matrix}. 
This method was used to simplify two loop SMEFT anomalous dimensions in \cite{Bern2020StructureMethods,EliasMiro:2020tdv}.
This relation between cuts and logs in scattering amplitudes was used in the past by the Russian school to calculate fixed order logs, see e.g.  \cite{Fadin:1995xg}.

However, 
the method in \cite{Caron-Huot2016RenormalizationS-matrix} does not address the issue of rapidity logs since it utilizes the variation
of matrix elements under scale transformations and, in the full theory, rapidity logs (such as in eq. (\ref{RRG})) are independent of $\mu$.
In this paper we show that one can extract the rapidity logs and their associated anomalous dimensions
by generalizing the ideas in  \cite{Caron-Huot2016RenormalizationS-matrix} and replacing the dilatations utilized in their derivation  by an special type of  complex boosts, once the amplitude is properly factorized into soft and collinear sectors.


At first sight it might seem strange to relate the phase of the S-matrix to the anomalous dimensions, as 
surely such a relation  cannot hold for a general process. If we consider a semi-classical approximation, for instance, the 
phase will correspond to the classical action 
\beq
{\cal M}_{SC} \sim e^{iS_{cl}},
\eeq
which is  not related to any RG anomalous dimension \footnote{The RG anomalous dimensions is associated with the hard quantum contribution, as discussed below.}.
However, for the canonical semi-classical scattering process of near forward, or eikonal scattering, the amplitude  is characterized by large  rapidity  logs of the form $\log(s/t)$, which are controlled
by the RAD, which in this case, is called the ``Regge trajectory" and is related to the phase.
In fact, it has been known for a long time that the Regge trajectory can be calculated from  the phase of
the S-matrix in planar Yang-Mills theory\cite{Kuraev:1976ge}, and moreover,  can be calculated exactly within the BDS ansatz \cite{Bern:2005iz} of $N=4$ SUSY \cite{Naculich:2007ub}. The phase contains both classical as well as quantum mechanical contributions, the
latter of which is related to the RAD. 
In this paper we will not be discussing the case of Regge Logs as they necessitate a slightly different formalism that
the one introduced here. Here will discuss the Regge trajectory calculation in a forthcoming publication \cite{RS2}.

In this paper we will be focusing on inherently quantum mechanics processes. That is processes for which there is a hard scattering.  We show how unitary can be used to extract RAD at the two loop level in two distinct cases.
The simplifications which arise using this phase/RAD relation are two-fold, as we will demonstrate:  It simplifies the integrals as  the phase arises due to the contribution from gluons whose
momentum resides in the ``Glauber region" where $k^\mu \sim (0,0,k_\perp)$ in light cone-coordinates.
Expanding around this region trivializes many of the integrals. Moreover, after expanding around
the Glauber region, the integrals are all finite in dimensional regularization, as no rapidity divergences arise. Thus, there is {\it no need to
introduce a rapidity regulator} that can complicate higher loop calculations. 

This paper is structured as follows. First, we review the results in \cite{Caron-Huot2016RenormalizationS-matrix} showing
how the phase of the (hard part) of the S-matrix can be used to extract the RG anomalous dimensions for observables. The generalization of these results to the case of rapidity anomalous dimensions follows once the factorization is proven, which is accomplished by invoking the Soft Collinear Effective Theory (SCET)\cite{Bauer:2001yt,Bauer:2001cu,Bauer:2002nz}. Once we have established a relationship between the
phase and the RAD, we illustrate the use of the formalism in two examples. First for a local operator, the Sudakov form factor and then for a non-local operator, the two parton transverse momentum soft function. 
To be complete we include an appendix which shows how the RAD anomalous dimensions are calculated using the 
canonical method following \cite{Chiu2012ATheory}.


We use the mostly minus metric $g_{\mu \nu}=Diag(1,-1,-1,-1)$ 
and define the usual $\overline{\text{MS}}$ factor
\begin{equation}
\bar{\mu}^{2\epsilon} = \mu^{2\epsilon} (4\pi)^{-\epsilon}e^{\epsilon \gamma_E},
\end{equation}
where we use $d=4-2 \epsilon$ and 
\beq
[d^dk]= \frac{d^dk}{(2 \pi)^d}.
\eeq
For the light cone coordinates we define two null vectors $n/\bar n^\mu=(1,0,0,\pm 1)$
and decompose four vectors as follows
\beq
p^\mu= n \cdot p \frac{\bar n^\mu}{2}+ \bar n \cdot p \frac{n^\mu}{2}+p_\perp^\mu \equiv (n\cdot p,\, \bar n\cdot p,\,p_\perp^\mu),
\eeq
such that 
\begin{equation}
p^-=n \cdot p, ~~ p^+=\bar n \cdot p .
\end{equation}
In this paper we will only be considering process with two ``jets", i.e. light like momenta which we will always take 
to be along the $\hat z$ axis. The generalization to more directions is straight-forward.

The phase space on shell delta function will be written as
\begin{equation}
\delta_+(p^2 - m^2) = \delta(p^2 - m^2) \bar \theta(\bar{n}\cdotp p + n\cdotp p),
\end{equation}
and $\bar \theta(x)\equiv (2 \pi) \theta(x)$. 

\section{The Master Formulae}

\subsection{The RG Master Formula}
 Reference \cite{Caron-Huot2016RenormalizationS-matrix} derives a relation between the S-matrix phase and
the RG anomalous dimension.
One starts by considering a generic form factor \footnote{Following the notation in \cite{Caron-Huot2016RenormalizationS-matrix}, any calligraphic character corresponds to a matrix element and not an operator.}
\beq
{\cal F}(p_1...p_n)= \int d^dx e^{-i q \cdot x} ~_{out}\langle p_1...p_n\mid F(x) \mid 0 \rangle_{in}
\eeq
where we will take all states outgoing in this way all of the invariants will be postive. This choice ensures that all the invariants are positive.
Consider the action of  the following complex dilatation ($D$) on this form factor
\beq
e^{i D \pi}{\cal F}(p_1...p_n)={\cal F}(-p_1...-p_n)=\int d^dx e^{-i q \cdot x} ~_{out}\langle 0 \mid F(x) \mid \bar p_1...\bar p_n \rangle_{in}
\eeq
where in the last line we utilized crossing symmetry and bar denotes the anti-particle. It follows that \beq
e^{i D \pi}{\cal F}(p_1...p_n)=\int d^dx e^{-i q \cdot x} ~_{in}\langle \bar p_1...\bar p_n \mid F^\star(x) \mid 0 \rangle_{out}^\star.
\eeq
Mechanically this transformation returns invariants back to their original form but now on the other side of the cut.

Then inserting $CPT(CPT)^{-1}$ appropriately into the matrix element we have
\beq
\label{res1}
e^{i D \pi}{\cal F}(p_1...p_n)={\cal F}(p_1...p_n)^\star.
\eeq

Next one treats $F$ as a perturbation to the S-matrix
\beq
S=S_0+iF
\eeq
such that the unitarity relation $SS^\dagger=1$ gives
\beq
S_0F^\dagger-FS_0^\dagger=0,
\eeq
where $F^2$ terms wont contribute to our matrix elements.
Then we have
\beq
\label{S}
F= S_0 F^\dagger S_0. 
\eeq
Restricting ourselves to the subset of matrix elements with no incoming particles we can effectively write
\beq
\label{old}
F= S_0 F^\dagger,
\eeq
and using (\ref{res1})
\beq
\label{newres}
e^{-i \pi D} {\cal F}^\star= {\cal S F^\star}.\eeq
Or in words: the anomalous dimensions equals the phase of the S-matrix. 
Now to make this result more user friendly
one writes the S-matrix element as ${\cal S}=1+i{\cal M}$. Expanding (\ref{newres}) to first order in perturbation theory \beq
\label{hard}
- \pi \gamma_\mu^{(1)} {\cal F}^{(0)\star}= {\cal M}^{(0)} {\cal F^\star}^{(1)}+{\cal M}^{(1)} {\cal F^\star}^{(0)},
\eeq
where $\gamma_\mu$ is the anomalous dimensions of $F$. Since the RHS corresponds to a matrix equation, 
we can consider any set of intermediate states between ${\cal M} $ and ${\cal F^\star}$.
As emphasized in \cite{Caron-Huot2016RenormalizationS-matrix}, this result needs to be refined due to the existence of ``IR anomalous
dimensions" ($\gamma_{\text{IR}}$), which corresponds to $\mu$ dependence introduced when regulating IR divergences.
So to extract $\gamma$ one must mod out by the appropriate matrix elements which
capture the IR divergences. It is important to emphasize that $\gamma_{\text{IR}}$ is not related
in any way to the rapidity anomalous dimensions. In the language of EFT, the relation (\ref{hard}) applies to the hard matching coefficients.

\subsection{Some Preliminaries about SCET}

We will be utilizing SCET for our proofs and to keep the paper self-contained
we will review the salient elements without going into details. 
The quantity of interest for this
paper is the rapidity anomalous dimension $\gamma_\nu$, which can be systematically defined
within the EFT as 
the analog to the canonical anomalous dimension and corresponds to the response of an operator to 
a change of the {\bf rapidity} factorization scale $\nu$ .  In the EFT one needs to introduce this scale to
distinguish between modes which have the same virtuality but hierarchaly different rapidity, as discussed below.
It is important to emphasize that once we have established  a connection between the rapidity anomalous dimension
and cut graphs, {\bf one need not calculate in the EFT}, instead if one prefers, one can calculate using the
full theory. The EFT integrands can be written down by
asymptotically expanding the full theory integrals into regions \cite{Smirnov:2002pj}.

The naive leading order action for SCET is written as 
\beq
S_{SCET}= S_{s}+S_n+S_{\bar n}
\eeq
Operators constituting power corrections may involve both types of fields, but since they are treated as perturbations,
their matrix elements will factorize into products of soft and collinear matrix elements. 
The collinear momenta scale as
\beq
k_n^\mu \sim(1,\lambda^2,\lambda)~~~~~~k_{\bar n}^\mu \sim(\lambda^2,1,\lambda)~~~~~~
\eeq
whereas the soft modes scale as
\beq
k_S^\mu \sim(\lambda,\lambda,\lambda),
\eeq
$\lambda$ is the power counting parameter which is given 
by the ratio of the UV to IR scales.

When this is the relevant set of modes the theory is called SCETII, as opposed to SCETI where the
soft modes (called ultra-soft) have virtuality $\lambda^4$. The fact that in SCETII, the soft and collinear modes
have the same virtuality leads to the need for a rapidity regulator, since they reside on the
same invariant mass hyperbola, and one needs to introduce a scale, which we call $\nu$, in order distinguish between them.

After integrating out the hard modes one generates as set of higher dimensional operators which mediate
any hard scattering. These operators are manifestly invariant under a separate set of gauge transformations, for
the soft and collinear sectors. For instance, for the Sudakov form factor, to be discussed below, we have the operator

\begin{equation}
F = \left(\bar{\xi}_n W_n\right) S_n^\dag\Gamma S_{\bar{n}} (W_{\bar{n}}^\dag \xi_{\bar{n}}),
\end{equation}
$W_n$ and $S_n$ are collinear and soft Wilson lines respectively, which ensure that the operator is invariant under separate 
gauge transformations.

What we have described here is an incomplete form of SCET since it does not include
the all-important Glauber modes which have momenta which scale as
\beq
k_g^\mu \sim( \lambda^2,\lambda^2,\lambda)
\eeq
in light cone coordinates. These modes are relevant as they can be exchanged
between collinear modes in opposite light cone directions, as well as between soft modes and collinear. 
In the original formulation of SCET this mode was assumed to either cancel
in observables of interest, or to not be necessary to reproduce the full theory amplitude, at least at leading
power \footnote{This was proven in the context of full QCD for a class of observables in \cite{Bodwin:1985ft,Collins:1984kg}.}. SCET was then completed to include these modes in \cite{Rothstein2016AnViolation}, where it was shown that
in a hard scattering process the Glaubers exchanged between ``active" partons (those which either enter or leave the hard scattering vertex) are actually included in soft exchanges \footnote{There maybe exceptions to this rule if the observable becomes sufficiently exclusive, in which case the Glaubers must be included into the theory and factorization breaks down.}. To properly handle this overlap of regions one may use the ``zero-bin" methodology \cite{Manohar:2006nz} whereby
one subtracts (in the case at hand) the Glauber limit of the soft integral and only then can one
consistently add the Glauber graph. 
The reader should consult section (10) of \cite{Rothstein2016AnViolation} for details.
The zero-bin limit of the soft graph is often identical to the Glauber graph. In which case one has a choice
to either ignore the Glauber graph (in the case of active-active interaction in a hard scattering process) 
or make the zero-bin subtraction and then add the Glauber. If one follows the former path, then the
direction of the Wilson line matters (i.e. to or from infinity), whereas performing the zero-bin subtraction
makes the direction irrelevant. As an example consider the one loop soft correction to a current
\beq
I_S\equiv \int \frac{d^dk}{(k^2+i \epsilon)} \frac{1}{(n \cdot k +i \epsilon)}\frac{1}{(\bar n \cdot k +i \epsilon)}
\eeq
Then the zero-bin subtraction of this graph is found by taking the Glauber limit of the soft moment, where $k_\pm \rightarrow 0$
\beq
I_S^{ZB}\equiv \int \frac{d^dk}{(-k_\perp^2+i \epsilon)} \frac{1}{(n \cdot k +i \epsilon)}\frac{1}{(\bar n \cdot k +i \epsilon)}
\eeq
which is identical to the Glauber graph. In this paper we will always calculate the Glauber contribution on its own 
and subtract zero bins when they are non-vanishing.

\subsection{The Glauber Operator(s)}

In this paper we will only be interested in processes where the Glauber is known to be absorbable into soft Wilson lines.
However, since the Glaubers are responsible for generating the phase we choosing to keep the Glaubers and implicitly zero-bin subtract the softs. The reason is that cuts will only come from the Glauber region since the Glaubers keep fermions on shell
whereas softs do not.

Operationally, in the EFT the Glaubers couple through the following operator
\beq
\label{Gbr}
O_g^{ns \bar n}= {\cal O}^A_n \frac{1}{{\cal{P}}_\perp^2} {\cal O}_s^{AB} \frac{1}{{\cal{P}}_\perp^2} {\cal O}_{\bar n}^B,
\eeq
where ${\cal P}_\perp$ is the transverse momentum derivative.
Here the superscript $ns\bar n$ is there to distinguish this operator, which involves all three sectors, from those
which involve only two sectors $(ns , \bar n s)$. 
The three operators ${\cal O}^A_n, {\cal O}^B_{\bar n}$ and ${\cal O}^{AB}_s$ can be found in \cite{Rothstein2016AnViolation} along
with the associated Feynman rules. The collinear operators can be bilinear in either quarks or gluons. The intermediate
operator ${\cal O}_s$ gives rise to the so-called Lipatov vertex which couples soft gluons to the t-channel Glauber exchange.
To make the operator consistent with soft and collinear gauge invariance one must insert the proper
set of Wilson lines which discussed in detail in section (6) of \cite{Rothstein2016AnViolation} .

Notice that this operator is leading order in the $\lambda$ power counting, due the enhancement of the factor of $1/k_\perp^2$, and
thus shatters factorization, at least in cases where their effects can not be absorbed into
soft Wilson lines, as is the case for the examples we will discuss in this paper.
The Glauber gluons
are responsible for generating the phase of the matrix associated with the
large rapidity logs since they are the modes which are sensitive to the pole structure.

\subsection{The Rapidity Renormalization Group}
As previously mentioned, when SCET is used to describe observables where the soft and collinear regions have the same
virtuality (SCETII), we introduce a rapidity factorization scale $\nu$, which also plays a role in regulating
rapidity divergences, in analogy with the RG scale $\mu$. Thus if we consider a general matrix element, whether local or not, it can always be written as
\beq
\label{fact}
\langle O \rangle= \langle O_s\rangle \otimes \langle O_n\rangle \otimes \langle O_{\bar n}\rangle
\eeq
where $\otimes$ represents a generalized convolution.
The individual pieces have rapidity divergences which are not regularized in dim. reg. Instead one regulates
them by introducing a rapidity regulator. Mechanically this is achieved by introducing a factor of
\beq
\label{reg}
(\frac{w^2 \mid n/\bar n \cdot {\cal P}\mid^{-\eta}}{\nu^{-\eta}},\frac{w \mid 2{\cal P}_z\mid^{-\eta/2}}{\nu^{-\eta/2}})
\eeq
for each collinear and soft eikonal Wilson line vertex, respectively. Notice the absolute value, as it will play an important role in what is to follow.
This can be implemented at the level of the action, and the reader is advised to consult \cite{Chiu2012ATheory} for details \footnote{Similar equations were written down in the CSS formalism but they were not based on a Lagrangian formalism \cite{Collins:1981zc}. }.
Higher order calculations involving hard scattering amplitudes have been accomplished using a different regulator \cite{Li:2016axz}.

The relative factors of two arise from the fact that the collinear gluons emissions are absorbed by Lagrangian vertices.
$w$ is a bookkeeping parameter which obeys 
\beq
\nu \frac{d}{d\nu}w= -\frac{\eta}{2} w
\eeq
and is set to one in the limit $\eta \rightarrow 0$.

This factorization in eq.(\ref{fact}) is not boost invariant, as such 
a transformation can slide soft and collinear regions into each other. However, the full matrix element is, of course, boost invariant.
Given a choice of frames, each matrix element depends upon the rapidity factorization scale
such that
\beq
\nu \frac{d}{d\nu} \langle O_i\rangle=\gamma^\nu_i \langle O_i\rangle,
\eeq
where $i$ includes both collinear and soft sectors (where $i=n,\bar n, s$).
Boost invariance then implies
\beq
\sum_i \gamma^\nu_i=0.
\eeq

In \cite{Rothstein2016AnViolation} a general formalism for deriving $\gamma_\nu$,  reviewed in the appendix, for a given process was presented
which shows how the $\nu$ dependence fixes the RAD, in analogy to traditional
RG anomalous dimensions calculations.
However, as we will see, the S-matrix methodology discussed above will simplify such calculations.

\subsection{The Rapidity Anomalous Dimensions Master Formula}

We now wish to  generalize the RG formalism  to the RRG case which follows 
once one uses the intuition gained from the work in \cite{Chiu2012ATheory} on the RRG.
In the case of RAD,  the relevant generator becomes $K_{z}$, the boost generator
in the $\hat z$ direction, instead of dilatations, $D$. However, it {\bf is not} the canonical boost in the following sense.
We notice from (\ref{RRG}) that if we want to move the singularity in the rapidity logs to the other side of the cut, as in the
case of the invariant mass logs,
we will need to boost the large $\pm$ light-cone momenta separately and independently, which is obviously not a
symmetry of the action.  This can be implemented by boosting each collinear sector separately:
\bea
p_n^\mu &= (p_n^+,\, p_n^-,\, p_\perp^\mu) \rightarrow (e^{ \gamma}p_n^+,\, e^{- \gamma}p_n^-,\, p_{n\perp}^\mu) \\
p_{\bn}^\mu  &= (p_{\bn}^+,\, p_{\bn}^-,\, p_{\bn\perp}^\mu) \rightarrow (e^{- \gamma}p_{\bn}^+,\, e^{ \gamma}p_{\bn}^-,\, p_{\bn\perp}^\mu).\nn
\eea
Using this operation we are able to transform $p_\mu \rightarrow -p_\mu$ by
choosing $\gamma =i\pi$, and by 
furthermore rotating by $\pi$ around the $\perp$ direction.  For each sector this rotation will act trivially on the
amplitude since each sector is invariant under rotations along the jet axis \footnote{We are not considering observables which may be sensitive to the angle between transverse momenta in differing jet directions, when there are more than two.}.
This modified transformation, whose generator we denote as $\bar K_z$, acts as follows
\begin{equation}
e^{i\pi \bar K_z}{\cal F}(p_1,...,p_n) ={\cal F}(-p_1,...,-p_n) ={\cal F}^\star(p_1,...,p_n),
\label{C.B.: conjugate}
\end{equation}
The modified boost generator acts only on the collinear sectors' momenta along the lightcone directions, so we may make the identification 
\beq
\bar K_z \equiv \sum_{i = n, \bn}K_z^i,
\eeq
where $K_z^i$ is the boost in the $i$th collinear sector's $z$-direction,
\bea
K_z^n = \sum_{\{p_j\in n\}}\left(p_j^+\frac{\partial}{\partial p_j^+} - p_j^-\frac{\partial}{\partial p_j^-}\right),\\
K_z^{\bn} =  \sum_{\{p_j\in \bn\}}\left(p_j^-\frac{\partial}{\partial p_j^-} - p_j^+\frac{\partial}{\partial p_j^+}\right).\nn
\eea
Dependence on  $p_\pm$ 
only appears in rapidity logarithms $\ln\mid p_i^\pm\mid /\nu$ in the collinear functions, where the absolute value follows from the definition of the
regulator.  Therefore, when operating on the
these functions we may make the replacement 
\beq
\bar K_z \simeq - \nu \frac{\partial}{\partial \nu}.
\eeq
Using the rapidity RGE equation
\begin{equation}
\nu\frac{\partial}{\partial \nu} O_{n/\bar n} = \gamma^{n/\bar n}_\nu O_{n/ \bar n},
\end{equation}
we have
\begin{equation}
{\bar K}_z\, O_{n/\bar{n}} = -\gamma^{n/\bar{n}}_\nu O_{n/\bar{n}}.
\end{equation}
Thus we can write
\bea
\label{RAD}
e^{i\pi {\bar K}_z}{\cal F} &=& \left(e^{i\pi {\bar K}_z} J_n\right)\otimes \mathcal{S}\otimes \left(e^{i\pi {\bar K}_z} J_{\bar{n}}\right),\nonumber\\
&=& e^{-i\pi(\gamma^n_\nu + \gamma^{\bar{n}}_\nu)}J_n\otimes \mathcal{S}\otimes J_{\bar{n}},\nonumber\\
&= &e^{i\pi\gamma^s_\nu} {\cal F}.
\eea
Now we use unitarity just as in the RG case to write
\beq
\label{above}
e^{-i\pi(\gamma^n_\nu + \gamma^{\bar{n}}_\nu)}{\cal F}^\star={\cal S} {\cal F}^\star=(1+i{\cal M}) {\cal F}^\star.
\eeq
It is worth going into the details of this result in the context of the effective theory. The rapidity regulator is defined
 such that the arguments of the logs involve $\mid p_\pm \mid$, and given that $p_\pm$ are defined to
be large, the action of $-p_\pm \frac{\partial}{\partial p_\pm}$ will simply yield $-\mid p_\pm \mid$.
Thus the action of the exponentiated generator will yield the phase, as shown in the previous equation, and thus $J(-p)\neq J^\star (p)$.
On the other hand, we know that the action on the entire form factor should result in a conjugation, implying that
the soft function contains a phase such that
\beq
J_n(-p) J_{\bar n}(-p) S= e^{i\pi\gamma^s_\nu}J_n(p) J_{\bar n}(p) S =( J_n(p) J_{\bar n}(p) S)^\star=( J_n(p) J_{\bar n}(p) S^\star),
\eeq
since $J$ is real.
We may conclude that
\beq
S^\star=e^{i\pi\gamma^s_\nu} S.
\eeq
This is a useful result since it means that we can calculate $\gamma_\nu$ using the master formula eq. (\ref{above}) by only considering
soft graphs, via the  replacement of $\cal{F}$ with $S$. Physically this result is a result of the fact that the 
phase of the amplitude comes from the soft region (determined by the direction of the Wilson line).

An obvious question which arises is, is it  the RG anomalous dimension or the RAD  which is related to the  phase of the
S-matrix? The answer is that the phase of the hard scattering piece gives the RG while the
phase of the IR piece gives the RAD \footnote{Since IR divergences are independent of $\nu$ they will not pose any obstruction.}. As previously emphasized since when we calculate we include
all of the modes, we could just as well calculate in the full theory, but if we do so we must
still subtract out the hard piece, which is typically a simpler calculation since it involves
integrals with fewer scales.

Another important distinction from the RG case is the fact that in SCET the imaginary part is divergent, 
as phase space is unbounded due to the use of the multipole expansion which is necessitated by the power counting \cite{Grinstein:1997gv}, and
we must amend eq.(\ref{newres}) to account for this fact.
Defining the renormalized operator 
\begin{equation}
F^{R} = Z_F^{-1}F^B,
\end{equation}
we have to revisit the result (\ref{old})
\beq
F= S_0 F^\dagger,
\eeq
which leads to \footnote{The distinction between $F$ and $\cal F$ become murky in the EFT as the
operators are tailored to the states in a very specific way.}
\beq
{\cal F}_B= {\cal S}_0 \frac{Z_F}{Z_F^\star} {\cal F_B}^\star.
\eeq
Then writing $S_0=1+i M$ we have 
\begin{equation}
\label{master}
e^{i\pi \gamma_\nu^s}{\cal F}^{\star B} =\frac{(Z_F^{\star})^{ -1}}{Z_F^{-1}}{\cal F}^{\star B}+ i\mathcal{M}{\cal F}^{\star B} \frac{(Z_F^{\star})^{ -1}}{Z_F^{-1}}.
\end{equation}
Expanded to one loop order, this gives
\begin{equation}
\label{RRGmasterloop1}
\gamma_\nu^{s(1)} {\cal F}^{\star(0)} = \frac{1}{\pi}\left((\mathcal{M} {\cal F}^\star)^{(1)} - 2 \text{Im}[Z_F^{-1}]^{(1)}{\cal F}^{\star(0)}\right).
\end{equation}



\subsection{Calculating in the Full Theory Versus the Effective Theory}\label{sec: 2.6}

The result (\ref{master})  was derived within the EFT where the hard part ($H$) had already be removed. 
To work in the full theory we must 
repristinate $H$ and allow for it to be complex. This is a trivial exercise with the result being that the new master formula is
\beq
e^{-i\pi\gamma^s_\nu}F^{R*} =\frac{H^\star}{H}S\otimes F^{R*},
\eeq
leaving
\beq
\label{fullt}
\gamma^s_\nu= \frac{i}{\pi}\log \left[ \frac{H^\star}{H}\left( 1+ i\sum_X  
\frac{ \langle \psi_f \mid X\rangle\langle X \mid  F^{R*} \mid 0 \rangle}{\langle \psi_f \mid   F^{R*} \mid 0 \rangle}
 \right) \right].
\eeq
This amendment to the EFT formula acts to remove any phase that might be generated by the hard part and is not
relevant to the RAD.
It is important to note that in this paper we will be doing all of our calculations in the effective theory. We have included
the result (\ref{fullt}) for those who would prefer to work in the full theory.

There are several advantages to working in the full theory. One can use the amplitudes tool box to skip having to write down Feynman diagrams. Furthermore, and perhaps more importantly, there may be   no need to introduce a rapidity regulator which can
lead to both nettlesome integrals as well as calculational subtleties.
At the same time it is also true that the effective theory  one need not regulate the rapidity divergences since we know that the RAD is finite.
However, to remove the rapidity regulator we would need to combine integrands coming from
various diagrams. The EFT calculation is also simplified by the fact that we only need to calculate soft graphs.

On the flip side, in the full theory  integrals are in general more difficult, though given the library of known integrals this may be an irrelevant fact. In the EFT one draws all possible Feynman diagrams,  of which, there can be many since
the theory is modal, i.e. split into regions. However, as we have discussed above, one need only concern oneself with the
soft sector.
In cases where the rapidity anomalous dimension is IR finite, we may simply ignore scaleless integrals. This is not the case when calculating RG anomalous dimensions where typically we would need to split such integrals into UV and IR pieces. Thus whether one chooses to work in the full or effective theory is a matter of convenience/taste.

\subsection{The structure of iterations}
Formula eq.(\ref{old}) has some interesting properties when we consider its expansion in the coupling as it  contains redundant information.  
Consider the expansion of eq.(\ref{old}) 
\beq
\label{one}
e^{i \gamma_s^\nu \pi}S^{R* (2)}= S^{R*(2)}+i( \gamma^{\nu(1)}_s( S^{R* (1)})+ \gamma^{\nu (2)}_s S^{R* (0)}) \pi- \frac{\pi^2}{2}(  \gamma^{\nu (1)}_s)^2S^{R*(0)}+...
\eeq
where here, for the sake of illustration we focus only on the terms which are second order in the coupling.
$S^{R*(n)}$ is the $n'th$ order contribution to the soft function. 
All of the terms aside from the one proportional to the target $\gamma^{\nu (2)}_s$ are Abelian in the sense that
they are scale as $C_F^2$ and can be considered  redundant information that need not be calculated. That
these terms cancel in the extraction of $\gamma^{\nu (2)}_s$ is a manifestation of non-Abelian exponentiation \cite{NAExpo1, NAExpo2}, 
which states, effectively, that the
sum of the  graphs for the product of any number of Wilson lines will exponentiate \cite{Gardi2013TheLines}  at the level of diagrams
where the only diagrams that contribute are those that are within a ``web" at a given order in the coupling. In the case of
two Wilson lines, as for the Sudakov form factor, a web consists of diagrams which are two particle eikonally irreducible in that they
can not be disconnected by cutting the two Wilson lines. At the order we are working  webs will always have color weights which
are linear in $C_F$  which is sometimes called``maximally-non-Abelian".  Anomalous dimensions for Wilson line operators
will always consist solely of webs, given that the coupling is independent of the rapidity scale $\nu$ and thus the solution to
the RRG is always a simple exponential.

 It is worthwhile to understand how the non-web pieces cancel in the result for $\gamma_s^{\nu(2)}$.
The RHS of the eq. (\ref{old}) has a second order contribution of the form
\beq
\label{two}
(\frac{Z_S^\star}{Z_S}{\cal S}\otimes S^{R*})_{(2)}= (\frac{Z_S^\star}{Z_S}(1+i M)\otimes S^{R*})_{(2)}=
 S^{R*(2)}+  (\frac{Z_S^\star}{Z_S})_{(1)}\ S^{R*(1)}+i M_{(1)}\otimes  S^{R*(1)}+iM_{(2)} S^{R*(0)}.
\eeq
Equating eq.(\ref{one}) and eq.(\ref{two}) we find
\beq
\label{form}
i( \gamma^{\nu(1)}_s( S^{R* (1)})+ \gamma^{\nu (2)}_s S^{R* (0)}) \pi- \frac{\pi^2}{2}(  \gamma^{\nu (1)}_s)^2S^{R*(0)}=i M_{(1)}\otimes  S^{R*(1)}+i(M_{(2)} S^{R*(0)})_{sub},
\eeq
where the counter-terms have been used to subtract the UV divergences from the final term on the RHS. Once we accept that
$\gamma^{\nu (2)}_s$ has a maximally non-Abelian structure, and we utilize the fact that  all of the other terms
on the LHS are NOT maximally non-Abelian, we may simply discard all of the $C_F^2$ pieces of the RHS to extract  $\gamma^{\nu (2)}_s$. 

An obvious question arises when one consider that the unitarity method, originally designed to calculate RG logs,
leads to an equation of the exact same form as eq. (\ref{fullt}). Why are the RG logs not also strictly given by webs?
The answer is that when one applies (\ref{fullt}) in the RG case it is applied to the hard piece only, whereas as we
are actually excising the hard piece from the full theory result.  Furthermore, non-Abelian exponentiation has
only been proven to apply to Wilson line observables, but our methods are  more general than that. In a forthcoming paper  \cite{RS2}    we will apply  our techniques to
 the case of forward scattering where the collinear lines do not eikonalize.

\section{The Sudakov Form Factor}

As our first example we consider Sudakov form factor which involves one IR (a mass) and one UV scale ($Q$) and
has the interesting property of containing double logarithms at each loop order which dominate its asymptotic
behavior. As usual to extract that anomalous dimensions we consider all out-going particles, i.e. we
will be considering this form factor in the time-like $Q^2>0$ region.
The double logs arise from overlapping soft and collinear divergences, and since the virtuality of the relevant modes (soft and collinear) 
are the same the result includes a rapidity divergence.
The resummation of these logs is an ancient subject (see for instance \cite{Collins:2011zzd}) upon which we hope to shed some new light 
as using our methodology can greatly simplify higher order computations. We will be considering two
distinct ways of representing the IR scale, by giving the gluon or quarks a mass. 
The massive gluon is in some sense the simpler case since the mass cuts off all IR divergences, but is
not terribly useful beyond one loop, since beyond that order we lose gauge invariance unless we are willing
to Higgs the theory. It would in fact be interesting to use the techniques introduced here to calculate
higher order corrections to the electroweak Sudakov form factor \cite{Electroweak1, Electroweak2}, but this goes beyond the scope of
this paper. On the other hand, using a quark mass has the advantage of maintaining gauge invariance at all orders but
needs dimensional regularization to handle the soft IR divergences. This form
factor is  not physical, since it is not IR safe, but in the case of the massive quark it is possible to generate a physical result by introducing a theory below the scale of the mass which will
absorb the IR divergences into long distance matrix elements \cite{Fleming:2007qr}.

\subsection{The Massive Gluon Sudakov Form Factor (MGFF)}

Let us now calculate the one loop value for $\gamma^S_\nu$ using our master formula eq.(\ref{RRGmasterloop1}). We are interested in the matrix element of the
soft  function that appears in the Sudakov form factor which is given by the product of two soft Wilson lines
\beq
S \equiv S_n S^\dagger_{\bar n}
\eeq
where in position space we have
\beq
S_n = P e^{i g \int_0^\infty d\lambda A(\lambda n)\cdot n d\lambda}. 
\eeq
Using our master formula we are interested in 
 the matrix element
\beq
\label{ME}
\langle p^n_1 p^{\bar n}_2\mid(\mathcal{M} S^\star)^{(1)} \mid 0 \rangle= \sum_X\langle p^n_1 p^{\bar n}_2 \mid \mathcal{M} \mid X \rangle \langle X \mid S^{\star{(1)}} \mid 0 \rangle,
\eeq
where, as previously discussed we need only concern ourselves with the soft piece to extract the RAD.
The intermediate state must involve both $n$ and $\bar n$ (eikonlized) partons, i.e.
\beq
\sum_X \rightarrow \sum_x \mid p^n_3 p_4^{\bar n} +x\rangle\langle p^n_3 p_4^{\bar n}+x \mid.
\eeq

As we have emphasized it is the Glauber region of the softs that generates the phase. So we include the Glauber operator in our action \footnote{The reader might be bothered by the fact that the Glauber operator includes fermionic fields and the soft function has not such field with which to contract (i.e. its a pure Wilson line). But this is just a technical misdirection as it is simple to just replace soft exchanges by Glauber exchanges.}
.
The soft contribution will not contribute to the phase once we perform the zero-bin subtraction.
Returning to eq.(\ref{RRGmasterloop1}), we see that there is only one diagram with a non-vanishing cut and it is given by
\begin{align}
\begin{gathered}
\scalebox{.6}{
\begin{tikzpicture}
\begin{feynman}
\vertex (a) at (0,1);
\vertex (c) at (0,-1);
\vertex [label = left: \(n\)] (e) at (-2, 1);
\vertex [label = left: \(\bar{n}\)] (f) at (-2, -1);
\vertex (g) at (-1, 1);
\vertex (h) at (-1, -1);
\node [crossed dot, NodeBlue] (b) at (1.41412, 0);
\vertex (c1) at (0,1.4);
\vertex (c2) at (0,-1.4);
\diagram* {
(f)--[charged scalar, line width= 0.3mm](h)--[charged scalar, line width= 0.3mm](c)--[charged scalar, line width= 0.3mm](b)--[charged scalar, line width= 0.3mm](a)--[charged scalar, line width= 0.3mm](g)--[charged scalar, line width= 0.3mm](e),
(c1)--[scalar, gray](c2),
(g)--[red,ghost, line width = 0.3mm](h)
};
\end{feynman}
\filldraw[red] (-1,1) ellipse (0.6mm and 1.2mm);
\filldraw[red] (-1,-1) ellipse (0.6mm and 1.2mm);
\end{tikzpicture}
}
\end{gathered}
= 2i \frac{\alpha_s}{\pi} C_F\,&\bar{\mu}^{2\epsilon}\int [d^d k]\frac{\,( \bar{n}\cdot p^\prime)\, n\cdot p}{k^2_\perp +M^2}\nonumber\\
&\times\delta_+(n \cdot p \bar n \cdot k - k_\perp^2) \delta_+(\bar n \cdot p^\prime n \cdot k + k_\perp^2).
\label{Sff: 1 loop}
\end{align}
We can also see that the $k_\perp$ integral is UV divergent. This is not too surprising, since as previously mentioned,  the phase-space integrals in the effective theory are often divergent due to the multipole expansion. 
Performing the integral, we find
\begin{align}
\begin{gathered}
\scalebox{.6}{
\begin{tikzpicture}
\begin{feynman}
\vertex (a) at (0,1);
\vertex (c) at (0,-1);
\vertex [label = left: \(n\)] (e) at (-2, 1);
\vertex [label = left: \(\bar{n}\)] (f) at (-2, -1);
\vertex (g) at (-1, 1);
\vertex (h) at (-1, -1);
\node [crossed dot, NodeBlue] (b) at (1.41412, 0);
\vertex (c1) at (0,1.4);
\vertex (c2) at (0,-1.4);
\diagram* {
(f)--[charged scalar, line width= 0.3mm](h)--[charged scalar, line width= 0.3mm](c)--[charged scalar, line width= 0.3mm](b)--[charged scalar, line width= 0.3mm](a)--[charged scalar, line width= 0.3mm](g)--[charged scalar, line width= 0.3mm](e),
(c1)--[scalar, gray](c2),
(g)--[red,ghost, line width = 0.3mm](h)
};
\end{feynman}
\filldraw[red] (-1,1) ellipse (0.6mm and 1.2mm);
\filldraw[red] (-1,-1) ellipse (0.6mm and 1.2mm);
\end{tikzpicture}
}
\end{gathered}
&= i C_F\alpha_s\, \bar{\mu}^{2\epsilon} (4\pi)^{\epsilon} \Gamma(\epsilon)M^{-2\epsilon}, \nonumber\\
& = i C_F\alpha_s\left[\frac{1}{\epsilon_{\text{UV}}} + \ln\frac{\mu^2}{M^2}\right].
\label{Sff1: cut}
\end{align}

We must also include the counter-term piece in eq. (\ref{RRGmasterloop1}), but, by construction, the imaginary part of the counter-term is nothing more than the negative of
the divergent part of eq.(\ref{ME}), so there is no need to perform any calculation, one may simply
drop the divergent part of eq.(\ref{ME}). Nonetheless as a check we may extract the counter-term from 
the full SCET (all sectors)   one loop correction to the current.
The  imaginary part of the Sudakov form factor soft function one-loop counter-term is given by
\beq
\text{Im}\l[ Z_F^{-1}\r]= -\frac{\alpha_s C_F}{2\epsilon_{\text{UV}}}.
\eeq
Using the master formula (\ref{RRGmasterloop1}) in conjunction with (\ref{Sff1: cut}), we obtain a finite result, and we deter-
mine the RAD to be
\begin{equation}
\gamma_\nu^s = \frac{C_F \alpha_s}{\pi}\ln\frac{M^2}{\mu^2}.
\end{equation}
Which agrees with the standard result for the soft RAD \cite{Chiu2012ATheory} for the massive gluon Sudakov form factor (see eq. (\ref{A1})). 

\subsection{The Massive Quark Sudakov Form Factor (MQFF)}

The inclusion of a quark mass avoids the gauge invariance issue, however at one loop the relevant integral is scaleless and technically vanishes.
However after separating and UV and IR divergences we can write
\begin{align}
\begin{gathered}
\scalebox{.6}{
\begin{tikzpicture}
\begin{feynman}
\vertex (a) at (0,1);
\vertex (c) at (0,-1);
\vertex [label = left: \(n\)] (e) at (-2, 1);
\vertex [label = left: \(\bar{n}\)] (f) at (-2, -1);
\vertex (g) at (-1, 1);
\vertex (h) at (-1, -1);
\node [crossed dot, NodeBlue] (b) at (1.41412, 0);
\vertex (c1) at (0,1.4);
\vertex (c2) at (0,-1.4);
\diagram* {
(f)--[charged scalar, line width= 0.3mm](h)--[charged scalar, line width= 0.3mm](c)--[charged scalar, line width= 0.3mm](b)--[charged scalar, line width= 0.3mm](a)--[charged scalar, line width= 0.3mm](g)--[charged scalar, line width= 0.3mm](e),
(c1)--[scalar, gray](c2),
(g)--[red,ghost, line width = 0.3mm](h)
};
\end{feynman}
\filldraw[red] (-1,1) ellipse (0.6mm and 1.2mm);
\filldraw[red] (-1,-1) ellipse (0.6mm and 1.2mm);
\end{tikzpicture}
}
\end{gathered}
&= i C_F\alpha_s\, \bar{\mu}^{2\epsilon} \int_0^\infty 
\frac{d^{2-2\epsilon}k_\perp}{k_\perp^2}
\nonumber\\
& = i C_F\alpha_s
(\frac{1}{\epsilon_{\text{UV}}}
- \frac{1}{\epsilon_{\text{IR}}} ).
\label{Sff: cut}
\end{align}
including the counter-term piece we find
\begin{equation}
\gamma_\nu^s = \frac{ \alpha_sC_F}{\pi}\frac{1}{\epsilon_{\text{IR}}},
\end{equation}
which agrees with result derived in the appendix using the canonical method (see eq.(\ref{canon})). 
As a check we can see that integrating the equation for the soft function
\beq
d S= \gamma^s_\nu S({d\ln\nu})
\eeq
by taking $S =1$ on the RHS we reproduce the term $\frac{\alpha_s C_F}{\pi}
\frac{1}{\epsilon_{\text{IR}}}\ln(Q/m)$
in the full theory result (see eq.(\ref{full})).

\subsection{Massive Quark Form-Factor at Two Loops}

The RAD  of the Sudakov form factor with massive matter lines has two incarnations. If there is no IR scales below the quark mass then
the form factor is an IR divergent quantity, as is its RAD. 
This is the case discussed in the previous section.
In more physical cases, there is an IR scale below the mass (typically
the QCD scale) and all IR divergences get absorbed into a non-perturbative low energy matrix element. 
This latter version, which we will focus on in this paper,  is relevant e.g., for resummations  in boosted top quark production and was first calculated in \cite{Hoang2015HardLoops}, where a  dispersive techniques was utilized  to circumvent the need to regulate the IR using a gluon mass which breaks gauge invariance beyond one loop. Here we demonstrate that we can extract the RAD by direct calculation by use of eq. (\ref{fullt}) without any reference to a gluon mass.  
 The one loop contribution shown in figure (1)  is  scaleless  and given by
\beq
\label{1loop}
(1)= -i\alpha_s C_F 
\left(\frac{1}{\epsilon_{UV}}-\frac{1}{\epsilon_{IR}} \right).
\eeq
The UV divergence will be killed by the counter-term for the current and the IR divergence will be dropped since
it will be factorized into a low energy matrix element. 
Thus  ($\gamma_s^{(1)\nu}=0)$.  As a consequence of this fact, all of the iterative terms in (\ref{form}) vanish.

Figure one shows all the diagrams that arise at $O(\alpha^2)$.  All of the diagrams 
are are scaleless save
for figure (1) where the massive quark is running through the loop.
\begin{figure}
\renewcommand\thesubfigure{\arabic{subfigure}}
\begin{subfigure}[b]{0.18\textwidth}
\caption{\qquad \qquad \qquad \qquad }
\centering
\scalebox{.7}{
\begin{tikzpicture}
\begin{feynman}
    \vertex (a) at (0,1);
    \vertex (c) at (0,-1);
    \vertex [label = left: \(n\)] (e) at (-2, 1);
    \vertex [label = left: \(\bar{n}\)] (f) at (-2, -1);
    \vertex (g) at (-1., 1);
    \vertex (h) at (-1., -1);
    \node [crossed dot, NodeBlue] (b) at (1.2, 0);
    \vertex (c1) at (0,1.4);
    \vertex (c2) at (0,-1.4);
    \vertex (s1) at (-1,.5);
    \vertex (s2) at (-1,-.5);
\diagram* {
    (f)--[charged scalar, line width= 0.3mm](h)--[charged scalar, line width= 0.3mm](c)--[charged scalar, line width= 0.3mm](b)--[charged scalar, line width= 0.3mm](a)--[charged scalar, line width= 0.3mm](g)--[charged scalar, line width= 0.3mm](e),
    (c1)--[scalar, gray](c2),
    (g)--[red,ghost, line width = 0.3mm](s1),
    (h)--[red,ghost, line width = 0.3mm](s2),
    (s1)--[half left, looseness = 1.75,SGreen,fermion, line width = 0.3mm](s2) --[half left, looseness = 1.75, SGreen,fermion, line width = 0.3mm](s1),
};
\end{feynman}
    \filldraw[red] (-1,1) ellipse (0.6mm and 1.2mm);
    \filldraw[red] (-1,-1) ellipse (0.6mm and 1.2mm);
    \filldraw[red] (-1,.5) ellipse (0.6mm and 1.2mm);
    \filldraw[red] (-1,-.5) ellipse (0.6mm and 1.2mm);
\end{tikzpicture}
}
\end{subfigure}
\begin{subfigure}[b]{0.18\textwidth}
\caption{\qquad \qquad \qquad \qquad }
\centering
\scalebox{.7}{
\begin{tikzpicture}
\begin{feynman}
    \vertex (a) at (0,1);
    \vertex (c) at (0,-1);
    \vertex [label = left: \(n\)] (e) at (-2, 1);
    \vertex [label = left: \(\bar{n}\)] (f) at (-2, -1);
    \vertex (g) at (-1.25, 1);
    \vertex (h) at (-1.25, -1);
    \node [crossed dot, NodeBlue] (b) at (1.2, 0);
    \vertex (c1) at (-0.5,1.4);
    \vertex (c2) at (-0.5,-1.4);
    \vertex (s1) at (.23, .6);
    \vertex (s2) at (.23, -.6);
\diagram* {
    (f)--[charged scalar, line width= 0.3mm](h)--[charged scalar, line width= 0.3mm](c)--[charged scalar, line width= 0.3mm](b)--[charged scalar, line width= 0.3mm](a)--[charged scalar, line width= 0.3mm](g)--[charged scalar, line width= 0.3mm](e),
    (c1)--[scalar, gray](c2),
    (g)--[red,ghost, line width = 0.9mm](h),
    (s2)--[SGreen, line width = 0.9](b)--[SGreen, line width = 0.9](s1)--[quarter right, SGreen, gluon, line width = 0.9](s2),
};
\end{feynman}
    \filldraw[red] (-1.25,1) ellipse (0.6mm and 1.2mm);
    \filldraw[red] (-1.25,-1) ellipse (0.6mm and 1.2mm);
\end{tikzpicture}
}
\end{subfigure}
\begin{subfigure}[b]{0.18\textwidth}
\caption{\qquad \qquad \qquad \qquad }
\centering
\scalebox{.7}{
\begin{tikzpicture}
\begin{feynman}
    \vertex (a) at (0,1);
    \vertex (c) at (0,-1);
    \vertex [label = left: \(n\)] (e) at (-2, 1);
    \vertex [label = left: \(\bar{n}\)] (f) at (-2, -1);
    \vertex (g) at (-1., 1);
    \vertex (h) at (-1., -1);
    \node [crossed dot, NodeBlue] (b) at (1.2, 0);
    \vertex (c1) at (0,1.4);
    \vertex (c2) at (0,-1.4);
    \vertex (s1) at (-1,.5);
    \vertex (s2) at (-1,-.5);
\diagram* {
    (f)--[charged scalar, line width= 0.3mm](h)--[charged scalar, line width= 0.3mm](c)--[charged scalar, line width= 0.3mm](b)--[charged scalar, line width= 0.3mm](a)--[charged scalar, line width= 0.3mm](g)--[charged scalar, line width= 0.3mm](e),
    (c1)--[scalar, gray](c2),
    (g)--[red,ghost, line width = 0.3mm](s1),
    (h)--[red,ghost, line width = 0.3mm](s2),
    (s1)--[half right, looseness = 1.75,SGreen, gluon,, line width = 0.3mm](s2) --[half right, looseness = 1.75, SGreen, gluon, line width = 0.3mm](s1),
};
\end{feynman}
    \filldraw[red] (-1,1) ellipse (0.6mm and 1.2mm);
    \filldraw[red] (-1,-1) ellipse (0.6mm and 1.2mm);
    \filldraw[red] (-1,.5) ellipse (0.6mm and 1.2mm);
    \filldraw[red] (-1,-.5) ellipse (0.6mm and 1.2mm);
\end{tikzpicture}
}
\end{subfigure}
\begin{subfigure}[b]{0.18\textwidth}
\caption{\qquad \qquad \qquad \qquad }
\centering
\scalebox{.7}{
\begin{tikzpicture}
\begin{feynman}
    \vertex (a) at (0,1);
    \vertex (c) at (0,-1);
    \vertex [label = left: \(n\)] (e) at (-2, 1);
    \vertex [label = left: \(\bar{n}\)] (f) at (-2, -1);
    \vertex (g) at (-1.5, 1);
    \vertex (h) at (-1.5, -1);
    \node [crossed dot, NodeBlue] (b) at (1.2, 0);
    \vertex (c1) at (0,1.4);
    \vertex (c2) at (0,-1.4);
    \vertex (s) at (-1.5,0);
\diagram* {
    (f)--[scalar, line width = 0.3mm](h)--[charged scalar, line width= 0.3mm](c)--[charged scalar, line width= 0.3mm](b)--[charged scalar, line width= 0.3mm](a)--[charged scalar, line width= 0.3mm](g)--[scalar, line width = 0.3mm](e),
    (c1)--[scalar, gray](c2),
    (g)--[red,ghost, line width = 0.3mm](h),
    (s)--[out=45, in=-45, loop, min distance=2cm, gluon, SGreen](s)
};
`    \draw (s) [gluon, SGreen, line width = 0.3mm]arc [start angle=-180, end angle=180, radius=0.5cm];
\end{feynman}
    \filldraw[red] (-1.5,1) ellipse (0.6mm and 1.2mm);
    \filldraw[red] (-1.5,-1) ellipse (0.6mm and 1.2mm);
    \filldraw[red] (-1.5,0) ellipse (0.6mm and 1.2mm);
\end{tikzpicture}
}
\end{subfigure}
\begin{subfigure}[b]{0.18\textwidth}
\caption{\qquad \qquad \qquad \qquad }
\centering
\scalebox{.7}{
\begin{tikzpicture}
\begin{feynman}
    \vertex (a) at (0,1);
    \vertex (c) at (0,-1);
    \vertex [label = left: \(n\)] (e) at (-2, 1);
    \vertex [label = left: \(\bar{n}\)] (f) at (-2, -1);
    \vertex (g) at (-1.5, 1);
    \vertex (h) at (-1.5, -1);
    \node [crossed dot, NodeBlue] (b) at (1.2, 0);
    \vertex (c1) at (0,1.4);
    \vertex (c2) at (0,-1.4);
    \vertex (s) at (-1.5,0);
\diagram* {
    (f)--[scalar, line width = 0.3mm](h)--[charged scalar, line width= 0.3mm](c)--[charged scalar, line width= 0.3mm](b)--[charged scalar, line width= 0.3mm](a)--[charged scalar, line width= 0.3mm](g)--[scalar, line width = 0.3mm](e),
    (c1)--[scalar, gray](c2),
    (g)--[red,ghost, line width = 0.3mm](h),
    (s)--[gluon, SGreen](b)
};
\end{feynman}
    \filldraw[red] (-1.5,1) ellipse (0.6mm and 1.2mm);
    \filldraw[red] (-1.5,-1) ellipse (0.6mm and 1.2mm);
    \filldraw[red] (-1.5,0) ellipse (0.6mm and 1.2mm);
\end{tikzpicture}
}
\end{subfigure}

\begin{subfigure}[b]{0.23\textwidth}
\caption{\qquad \qquad \qquad }
\centering
\scalebox{.7}{
\begin{tikzpicture}
\begin{feynman}
    \vertex (a) at (0,.75);
    \vertex (c) at (0,-.75);
    \vertex [label = left: \(n\)] (e) at (-2, .75);
    \vertex [label = left: \(\bar{n}\)] (f) at (-2, -.75);
    \vertex (g) at (-1, .75);
    \vertex (h) at (-1, -.75);
    \node [crossed dot, NodeBlue] (b) at (1.2, 0);
    \vertex (c1) at (0,1.4);
    \vertex (c2) at (0,-1.4);
    \vertex [label = right:\(Z_\alpha\)] (s) at (-1,0);
\diagram* {
    (f)--[charged scalar, line width= 0.3mm](h)--[charged scalar, line width= 0.3mm](c)--[charged scalar, line width= 0.3mm](b)--[charged scalar, line width= 0.3mm](a)--[charged scalar, line width= 0.3mm](g)--[charged scalar, line width= 0.3mm](e),
    (c1)--[scalar, gray](c2),
    (g)--[red,ghost, line width = 0.3mm](h)
};
\end{feynman}
    \filldraw[red] (-1,.75) ellipse (0.6mm and 1.2mm);
    \filldraw[red] (-1,-.75) ellipse (0.6mm and 1.2mm);
    \filldraw[red] (-1,0) ellipse (0.6mm and 1.2mm);
\end{tikzpicture}
}
\end{subfigure}
\begin{subfigure}[b]{0.23\textwidth}
\caption{\qquad \qquad \qquad \qquad }
\centering
\scalebox{.7}{
\begin{tikzpicture}
\begin{feynman}
    \vertex (a) at (0,.75);
    \vertex (c) at (0,-.75);
    \vertex [label = left: \(n\)] (e) at (-2.5, .75);
    \vertex [label = left: \(\bar{n}\)] (f) at (-2.5, -.75);
    \vertex (g1) at (-.5, .75);
    \vertex (h1) at (-.5, -.75);
    \vertex (g2) at (-2, .75);
    \vertex (h2) at (-2, -.75);
    \node [crossed dot, NodeBlue] (b) at (1.2, 0);
    \vertex (c1) at (0,1.4);
    \vertex (c2) at (0,-1.4);
\diagram* {
    (f)--[charged scalar, line width= 0.3mm](c)--[charged scalar, line width= 0.3mm](b)--[charged scalar, line width= 0.3mm](a)--[charged scalar, line width= 0.3mm](e),
    (c1)--[scalar, gray](c2),
    (g1)--[red,ghost, line width = 0.3mm](h1),
    (g2)--[red,ghost, line width = 0.3mm](h2)
};
\end{feynman}
    \filldraw[red] (-2,.75) ellipse (0.6mm and 1.2mm);
    \filldraw[red] (-2,-.75) ellipse (0.6mm and 1.2mm);
    \filldraw[red] (-.5,.75) ellipse (0.6mm and 1.2mm);
    \filldraw[red] (-.5,-.75) ellipse (0.6mm and 1.2mm);
\end{tikzpicture}
}
\end{subfigure}
\begin{subfigure}[b]{0.23\textwidth}
\caption{\qquad \qquad \qquad \qquad }
\centering
\scalebox{.7}{
\begin{tikzpicture}
\begin{feynman}
    \vertex (a) at (0,.75);
    \vertex (c) at (0,-.75);
    \vertex [label = left: \(n\)] (e) at (-2.5, .75);
    \vertex [label = left: \(\bar{n}\)] (f) at (-2.5, -.75);
    \vertex (g1) at (-.5, .75);
    \vertex (h1) at (-.5, -.75);
    \vertex (g2) at (-2, .75);
    \vertex (h2) at (-2, -.75);
    \node [crossed dot, NodeBlue] (b) at (1.2, 0);
    \vertex (c1) at (-1.25,1.4);
    \vertex (c2) at (-1.25,-1.4);
\diagram* {
    (f)--[scalar, line width= 0.3mm](c)--[charged scalar, line width= 0.3mm](b)--[charged scalar, line width= 0.3mm](a)--[scalar, line width= 0.3mm](e),
    (c1)--[scalar, gray](c2),
    (g1)--[red,ghost, line width = 0.3mm](h1),
    (g2)--[red,ghost, line width = 0.3mm](h2)
};
\end{feynman}
    \filldraw[red] (-2,.75) ellipse (0.6mm and 1.2mm);
    \filldraw[red] (-2,-.75) ellipse (0.6mm and 1.2mm);
    \filldraw[red] (-.5,.75) ellipse (0.6mm and 1.2mm);
    \filldraw[red] (-.5,-.75) ellipse (0.6mm and 1.2mm);
\end{tikzpicture}
}
\end{subfigure}
\begin{subfigure}[b]{0.23\textwidth}
\caption{\qquad \qquad \qquad \qquad }
\centering
\scalebox{.7}{
\begin{tikzpicture}
\begin{feynman}
    \vertex (a) at (0,.75);
    \vertex (c) at (0,-.75);
    \vertex [label = left: \(n\)] (e) at (-2.5, .75);
    \vertex [label = left: \(\bar{n}\)] (f) at (-2.5, -.75);
    \vertex (g1) at (-.5, .75);
    \vertex (h1) at (-.5, -.75);
    \vertex (g2) at (-2, .75);
    \vertex (h2) at (-2, -.75);
    \node [crossed dot, NodeBlue] (b) at (1.2, 0);
    \vertex (c1) at (0,1.4);
    \vertex (c2) at (0,-1.4);
\diagram* {
    (f)--[charged scalar, line width= 0.3mm](c)--[charged scalar, line width= 0.3mm](b)--[charged scalar, line width= 0.3mm](a)--[charged scalar, line width= 0.3mm](e),
    (c1)--[scalar, gray](c2),
    (g1)--[red,ghost, line width = 0.3mm](h2),
    (g2)--[red,ghost, line width = 0.3mm](h1)
};
\end{feynman}
    \filldraw[red] (-2,.75) ellipse (0.6mm and 1.2mm);
    \filldraw[red] (-2,-.75) ellipse (0.6mm and 1.2mm);
    \filldraw[red] (-.5,.75) ellipse (0.6mm and 1.2mm);
    \filldraw[red] (-.5,-.75) ellipse (0.6mm and 1.2mm);
\end{tikzpicture}
}
\end{subfigure}
\caption{Cut diagrams contributing to the 2-loop Sudakov RAD. 
Greens lines are soft (zero-bin subtracted) while dotted are Glauber. All of the diagrams are scaleless except for the first.}
\label{fig: 3}
\end{figure}
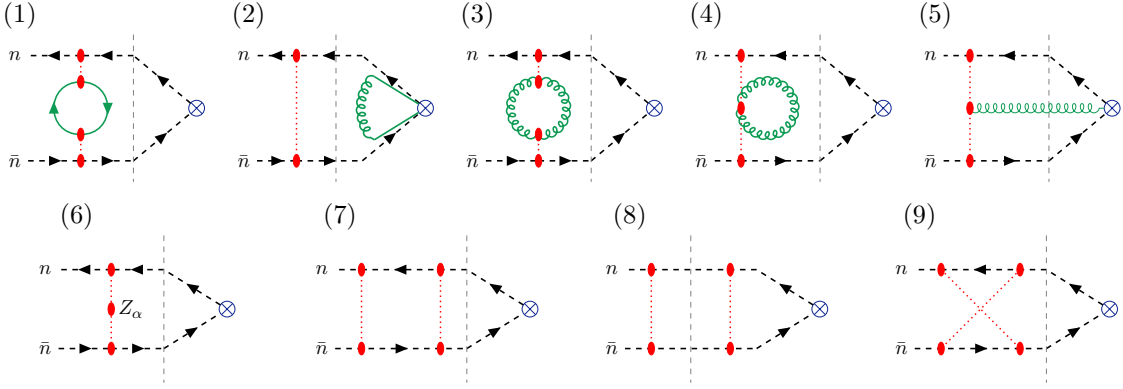
We conclude that we can read off the anomalous dimension at two loops from  diagram, 
\begin{align}
\begin{gathered}
\scalebox{.5}{
\begin{tikzpicture}
\begin{feynman}
    \vertex (a) at (0,1);
    \vertex (c) at (0,-1);
    \vertex [label = left: \(n\)] (e) at (-2, 1);
    \vertex [label = left: \(\bar{n}\)] (f) at (-2, -1);
    \vertex (g) at (-1., 1);
    \vertex (h) at (-1., -1);
    \node [crossed dot, NodeBlue] (b) at (1.2, 0);
    \vertex (c1) at (0,1.4);
    \vertex (c2) at (0,-1.4);
    \vertex (s1) at (-1,.5);
    \vertex (s2) at (-1,-.5);
\diagram* {
    (f)--[charged scalar, line width= 0.3mm](h)--[charged scalar, line width= 0.3mm](c)--[charged scalar, line width= 0.3mm](b)--[charged scalar, line width= 0.3mm](a)--[charged scalar, line width= 0.3mm](g)--[charged scalar, line width= 0.3mm](e),
    (c1)--[scalar, gray](c2),
    (g)--[red,ghost, line width = 0.3mm](s1),
    (h)--[red,ghost, line width = 0.3mm](s2),
    (s1)--[half left, looseness = 1.75,SGreen,fermion, line width = 0.3mm](s2) --[half left, looseness = 1.75, SGreen,fermion, line width = 0.3mm](s1),
};
\end{feynman}
    \filldraw[red] (-1,1) ellipse (0.6mm and 1.2mm);
    \filldraw[red] (-1,-1) ellipse (0.6mm and 1.2mm);
    \filldraw[red] (-1,.5) ellipse (0.6mm and 1.2mm);
    \filldraw[red] (-1,-.5) ellipse (0.6mm and 1.2mm);
\end{tikzpicture}
}
\end{gathered}
&= 64\pi^2 C_F\,T_F\,S_\Gamma \int \frac{\bar{\mu}^{4\varepsilon}\dbar^{d-2}k_\perp \dbar^d q}{(q^2-m^2)\,((q + k_\perp)^2-m^2)\, (k_\perp^2)^2}\text{tr}\left[(\slashed{q} + \slashed{k}_\perp + m)\frac{\slashed{n}}{2}(\slashed{q}-m)\frac{\slashed{\bar{n}}}{2}\right],\nonumber\\
&=i\frac{C_F\,T_F\,\alpha_s^2 }{\pi} \frac{\Gamma(2\varepsilon)\Gamma(2 + 2\varepsilon)^2}{\varepsilon\,\Gamma(4  +2\varepsilon)}\left(\frac{\mu^2}{m^2}\right)^{2\varepsilon},\\
& = i \pi\, C_F\, T_F\,\frac{\alpha_s^2}{\pi}\left(-\frac{1}{6\varepsilon^2} - \frac{1}{3\varepsilon}\log\frac{\mu^2}{m^2} + \frac{5}{18\varepsilon} - \frac{1}{3}\log^2\frac{\mu^2}{m^2} + \frac{5}{9}\log\frac{\mu^2}{m^2} - \frac{\pi^2}{36} - \frac{14}{27}\right).\nonumber
\label{soft vacuum pol.: full}
\end{align}
This integral can be evaluated by evaluating the $q$-integal first before the $k_\perp$-integral; this is recursively one-loop and can be handled by, e.g. Feynman parameters. 

We can extract that RAD by dropping all the poles, since the counter-term contribution assures us of a 
UV finite result, and the previous argument above  permits us to drop the IR divergent pieces. 
Thus we are left with 
\beq
\gamma^s_\nu=C_F\, T_F\,\frac{\alpha_s^2}{\pi^2}\left( - \frac{1}{3}\log^2\frac{\mu^2}{m^2} + \frac{5}{9}\log\frac{\mu^2}{m^2} - \frac{\pi^2}{36} - \frac{14}{27}\right).
\eeq

By keeping the quark masses in the loops and using $\overline{MS}$ we are inherently  working in 
a  non-decoupling (ND) scheme.  It is therefore prudent to change back to the usual EFT/$\overline{MS}$ scheme
where the quarks are taken to be massless. 
The relation between these schemes is
\beq
\alpha_s^{ND}=\alpha_s(1+ (\Pi(m^2,0)-\frac{\alpha_s}{3 \pi} \frac{T_F}{\epsilon})).
\eeq
where $\Pi$ is the scalar part of the vacuum polarization at $k=0$. We must make this replacement in the one loop 
IR divergent result and subsequently drop all the IR divergent terms. Thus the results will only be sensitive
to the $O(\epsilon)$ piece of $\Pi$
\begin{equation}
\Pi(m^2,0)_\epsilon = \frac{\alpha T_F}{3 \pi}\epsilon\left( \frac{1}{2}\log^2\frac{\mu^2}{m^2} + \frac{\pi^2}{12}\right),
\end{equation}
The net sum gives for the RAD
\beq
\gamma^s_\nu=-C_F\, T_F\,\frac{\alpha_s^2}{3\pi^2}\left(\frac{1}{2}\log^2\frac{m^2}{\mu^2}+ \frac{5}{3}\log\frac{m^2}{\mu^2} + \frac{14}{9}\right)
\eeq
 which agrees with the result given  in \cite{Hoang2015HardLoops}.

\section{Form Factors of non-Local Operators: The Soft Function}

Next we will apply the unitarity technique to calculate the RAD of a non-local operator. In particular we will consider the soft function  that arises in 
the factorization of a class of hard scattering observables. We will choose one particular operator
but the method can be applied more generally. We are interested in operators whose matrix elements
include rapidity divergences/logs. As such, we will choose observables whose diagrams include
soft and collinear modes of the same virtuality, the classic example of which are differential cross sections
where one measures the transverse momentum ($p_\perp$) of some set of particles, with $p_\perp\ll Q$, where $Q$ is the hard scattering scale.
Schematically the cross section takes the form
\beq
\frac{d\sigma}{dp_\perp}= H(Q,\mu) C_n(p_\perp/\mu,n \cdot p/\nu) \otimes C_{\bar n}(p_\perp/\mu,\bar n \cdot p^\prime/\nu)\otimes S(\mu/\nu,p_\perp/\mu ),
\eeq
where $\otimes$ denotes a convolution in the 
momentum variables. $n \cdot p$ and $\bar n \cdot p^\prime$ are 
the incoming light-cone momenta, which are integrated over weighted by PDF's.
The collinear pieces are transverse momentum parton distribution functions (TMPDFs) while
the soft function is the vacuum expectation value of Soft Wilson lines in the fundamental representation
\begin{equation}
S(b_\perp) =\frac{1}{N_c}\langle 0 \mid S^{ce}_n( b_\perp; 0,\infty) S^{\dagger ed}_{\bar n} (b_\perp;0,-\infty) S^{de}_{\bar n}( 0;0, -\infty)S^{\dagger ec} _{ n}(0; 0,\infty)
\mid 0 \rangle,
\end{equation}
where for convenience we have Fourier transformed to impact parameter space.
One can also consider double, or higher, order $n$-parton scattering in which case the soft function 
becomes the non-local product of $n$ pairs of Wilson lines with each in a different light cone-direction.
The arguments below are easily generalizable beyond the two parton scattering we consider here.
 In trying to use our unitarity methods however, we are immediately met with the fact that
$S$ is a  real valued function.
We can circumvent this problem by looking at a different matrix element which shares the same RAD.
In analogy with what we did for the Sudakov form factor we will consider particle production with
four outgoing states. 
Recall that this step was also necessary to eliminate the factor of $S$ on the left-hand side of
eq.(\ref{S}).
A similar calculation was done for the RG anomalous dimensions for parton distributions in \cite{Caron-Huot2016RenormalizationS-matrix}.



We will consider the matrix element with all out going partons
\beq
M(b_\perp)=\sum_X \langle p_n \bar p_{\bar n} q_n \bar q_{\bar n}\mid X \rangle \langle X \mid \bar S(b) \mid 0 \rangle.
\eeq
The ``crossed'' Wilson line is then
\beq
\bar S(b_\perp) =\frac{1}{N_c}\langle 0 \mid S^{ce}_n( b_\perp; \infty,0) S^{\dagger ed}_{\bar n} (b_\perp ;0,-\infty) S^{df}_n(0; \infty,0) S^{\dagger fc}_{\bar n} (0;0,-\infty).
\mid 0 \rangle,
\end{equation}

We have dropped the dependence on $x_\pm$ since the RAD cannot depend
upon the parton light-cone momentum fraction. 
At one loop there are only two diagrams.
The diagrams where the Glauber connects partons with the
same impact parameter will be scaleless  and can be dropped. Furthermore, there are no soft exchanges
since they have no cut piece (recall the phase comes from the Glauber piece of the soft)  except for self-energy diagrams which vanish. Thus
at one loop we have two diagrams which give the identical result:
\begin{align}
\begin{gathered}
\scalebox{.5}{
\begin{tikzpicture}
\begin{feynman}
\node [crossed dot, NodeBlue] (b1) at (1.5, .5);
\node [crossed dot, NodeBlue] (b2) at (1.5, -.5);
\node at (2,0) {$b_\perp$};
\vertex (a1) at (0, 1.5);
\vertex (c1) at (0,-0.75);
\vertex (a2) at (0, .75);
\vertex (c2) at (0,-1.5);
\vertex (e1) at (0.99,-0.075);
\vertex (e2) at (0.81, 0.075);
\vertex (cc1) at (0, 2);
\vertex (cc2) at (0,-2);
\vertex (n1) at (-2, 1.5);
\vertex (n2) at (-2, 0.75);
\vertex (nb2) at (-2, -1.5);
\vertex (nb1) at (-2, -0.75);
\node at (-2,-1.125) {$\bar{n}$};
\node at (-2,1.125) {$n$};
\vertex (g1) at (-1, .75);
\vertex (g2) at (-1, -.75);
\diagram* {
(n2)--[ line width = 0.2mm](a2)--[ line width = 0.2mm](e2),
(e1)--[ line width = 0.2mm](b2)--[ line width = 0.2mm](c2)--[ line width = 0.2mm](nb2),
(b1)--[scalar](b2),
(cc1)--[scalar, gray](cc2),
(g1)--[scalar, red, line width = 0.3mm](g2),
(n1)--[ line width = 0.2mm](a1)--[ line width = 0.2mm](b1)--[ line width = 0.2mm](c1)--[ line width = 0.2mm](nb1),
};
\end{feynman}
\filldraw[red] (-1,.75) ellipse (0.6mm and 1.2mm);
\filldraw[red] (-1,-.75) ellipse (0.6mm and 1.2mm);
\end{tikzpicture}
}
\end{gathered}
+
\begin{gathered}
\scalebox{.5}{
\begin{tikzpicture}
\begin{feynman}
\node [crossed dot, NodeBlue] (b1) at (1.5, .5);
\node [crossed dot, NodeBlue] (b2) at (1.5, -.5);
\node at (2,0) {$b_\perp$};
\vertex (a1) at (0, 1.5);
\vertex (c1) at (0,-0.75);
\vertex (a2) at (0, .75);
\vertex (c2) at (0,-1.5);
\vertex (e1) at (0.99,-0.075);
\vertex (e2) at (0.81, 0.075);
\vertex (cc1) at (0, 2);
\vertex (cc2) at (0,-2);
\vertex (n1) at (-2, 1.5);
\vertex (n2) at (-2, 0.75);
\vertex (nb2) at (-2, -1.5);
\vertex (nb1) at (-2, -0.75);
\node at (-2,-1.125) {$\bar{n}$};
\node at (-2,1.125) {$n$};
\vertex (g1) at (-1, 1.5);
\vertex (g2) at (-1, -1.5);
\diagram* {
(n2)--[ line width = 0.2mm](a2)--[ line width = 0.2mm](e2),
(e1)--[ line width = 0.2mm](b2)--[ line width = 0.2mm](c2)--[ line width = 0.2mm](nb2),
(b1)--[scalar](b2),
(cc1)--[scalar, gray](cc2),
(g1)--[scalar, red, line width = 0.3mm](g2),
(n1)--[ line width = 0.2mm](a1)--[ line width = 0.2mm](b1)--[ line width = 0.2mm](c1)--[ line width = 0.2mm](nb1),
};
\end{feynman}
\filldraw[red] (-1,1.5) ellipse (0.6mm and 1.2mm);
\filldraw[red] (-1,-1.5) ellipse (0.6mm and 1.2mm);
\end{tikzpicture}
}
\end{gathered}
&= 2i\,g^2 C_F \int\frac{ \dbar^{d-2}k_\perp}{k_\perp^2} e^{-ik_\perp\cdotp b_\perp}\nonumber\\
&= 2i\,g^2 C_F\frac{\Gamma(-\epsilon)}{4\pi}(\tilde{b}^2\mu^2)^{\epsilon}\\
&=-C_F \frac{i\,g^2}{2\pi}\left(\frac{1}{\epsilon} + \ln (\tilde{b}^2\mu^2) + O(\epsilon)\right),\nonumber
\label{TMD}
\end{align}
with $\tilde{b}^2 = b_\perp^{\,2} e^{\gamma_E}/4$. Dropping the UV divergence, we have exactly the one-loop RAD
\begin{equation}
\gamma_\nu^{s(1)} = \frac{\alpha_s}{\pi}(2 C_F)\ln (\tilde{b}^2\mu^2).
\end{equation}

\subsection{TMD Two-Loop Rapidity Anomalous Dimension}

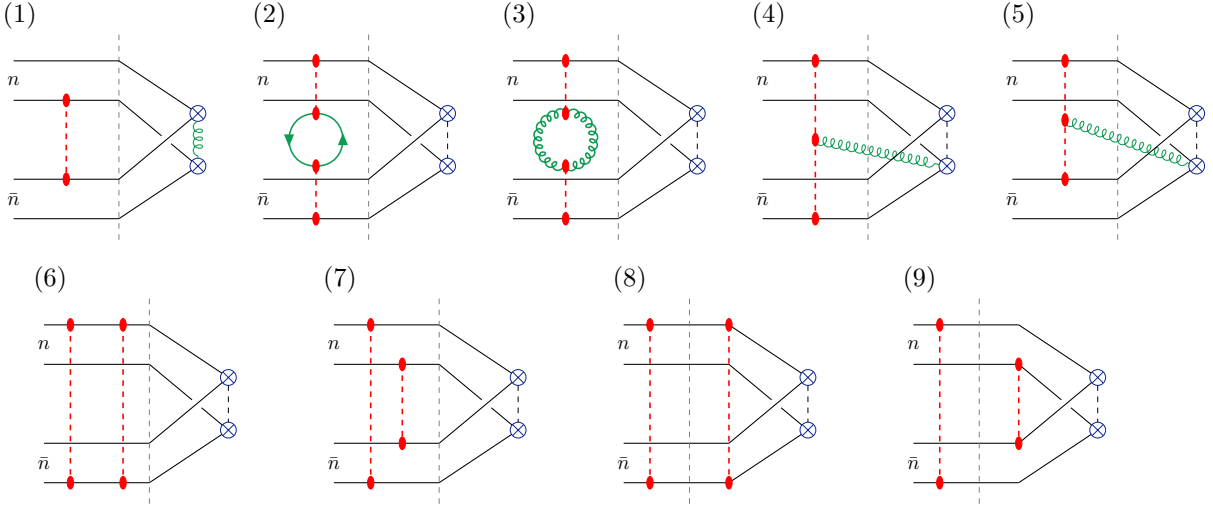
\begin{figure}
\renewcommand\thesubfigure{\arabic{subfigure}}
\begin{subfigure}[b]{0.18\textwidth}
\caption{\qquad \qquad \qquad \qquad }
\centering
\scalebox{.7}{
\begin{tikzpicture}
\begin{feynman}
    \node [crossed dot, NodeBlue] (b1) at (1.5, .5);
    \node [crossed dot, NodeBlue] (b2) at (1.5, -.5);
	\vertex (a1) at (0, 1.5);
	\vertex (c1) at (0,-0.75);
	\vertex (a2) at (0, .75);
	\vertex (c2) at (0,-1.5);
	\vertex (e1) at (0.99,-0.075);
	\vertex (e2) at (0.81, 0.075);
	\vertex (cc1) at (0, 2);
	\vertex (cc2) at (0,-2);
	\vertex  (n1) at (-2, 1.5);
	\vertex  (n2) at (-2, 0.75);
	\vertex  (nb2) at (-2, -1.5);
	\vertex  (nb1) at (-2, -0.75);
	\node at (-2,-1.125) {$\bar{n}$};
	\node at (-2,1.125) {$n$};
	\vertex (g1) at (-1, .75);
	\vertex (g2) at (-1, -.75);
\diagram* {

(n2)--[ line width = 0.2mm](a2)--[ line width = 0.2mm](e2),
(e1)--[ line width = 0.2mm](b2)--[ line width = 0.2mm](c2)--[ line width = 0.2mm](nb2),
(b1)--[gluon, SGreen, line width = 0.2mm](b2),
(cc1)--[scalar, gray](cc2),
(g1)--[scalar, red, line width = 0.3mm](g2),
(n1)--[ line width = 0.2mm](a1)--[ line width = 0.2mm](b1)--[ line width = 0.2mm](c1)--[ line width = 0.2mm](nb1),
};
\end{feynman}
    \filldraw[red] (-1,.75) ellipse (0.6mm and 1.2mm);
    \filldraw[red] (-1,-.75) ellipse (0.6mm and 1.2mm);
\end{tikzpicture}
}
\end{subfigure}
\begin{subfigure}[b]{0.18\textwidth}
\caption{\qquad \qquad \qquad \qquad }
\centering
\scalebox{.7}{
\begin{tikzpicture}
\begin{feynman}
    \node [crossed dot, NodeBlue] (b1) at (1.5, .5);
    \node [crossed dot, NodeBlue] (b2) at (1.5, -.5);
	\vertex (a1) at (0, 1.5);
	\vertex (c1) at (0,-0.75);
	\vertex (a2) at (0, .75);
	\vertex (c2) at (0,-1.5);
	\vertex (e1) at (0.99,-0.075);
	\vertex (e2) at (0.81, 0.075);
	\vertex (cc1) at (0, 2);
	\vertex (cc2) at (0,-2);
	\vertex  (n1) at (-2, 1.5);
	\vertex  (n2) at (-2, 0.75);
	\vertex  (nb2) at (-2, -1.5);
	\vertex  (nb1) at (-2, -0.75);
	\node at (-2,-1.125) {$\bar{n}$};
	\node at (-2,1.125) {$n$};
	\vertex (g1) at (-1, 1.5);
	\vertex (g2) at (-1, -1.5);
	\vertex (s1) at (-1, .5);
	\vertex (s2) at (-1,-.5);
\diagram* {

(n2)--[ line width = 0.2mm](a2)--[ line width = 0.2mm](e2),
(e1)--[ line width = 0.2mm](b2)--[ line width = 0.2mm](c2)--[ line width = 0.2mm](nb2),
(b1)--[scalar](b2),
(cc1)--[scalar, gray](cc2),
(g1)--[scalar, red, line width = 0.3mm](s1),
(g2)--[scalar, red, line width = 0.3mm](s2),
(s1)--[half right, looseness = 1.75,SGreen, fermion, line width = 0.3mm](s2) --[half right, looseness = 1.75, SGreen, fermion, line width = 0.3mm](s1),
(n1)--[ line width = 0.2mm](a1)--[ line width = 0.2mm](b1)--[ line width = 0.2mm](c1)--[ line width = 0.2mm](nb1),
};
\end{feynman}
    \filldraw[red] (-1,1.5) ellipse (0.6mm and 1.2mm);
    \filldraw[red] (-1,-1.5) ellipse (0.6mm and 1.2mm);
    \filldraw[red] (-1,.5) ellipse (0.6mm and 1.2mm);
    \filldraw[red] (-1,-.5) ellipse (0.6mm and 1.2mm);
\end{tikzpicture}
}
\end{subfigure}
\begin{subfigure}[b]{0.18\textwidth}
\caption{\qquad \qquad \qquad \qquad }
\centering
\scalebox{.7}{
\begin{tikzpicture}
\begin{feynman}
    \node [crossed dot, NodeBlue] (b1) at (1.5, .5);
    \node [crossed dot, NodeBlue] (b2) at (1.5, -.5);
	\vertex (a1) at (0, 1.5);
	\vertex (c1) at (0,-0.75);
	\vertex (a2) at (0, .75);
	\vertex (c2) at (0,-1.5);
	\vertex (e1) at (0.99,-0.075);
	\vertex (e2) at (0.81, 0.075);
	\vertex (cc1) at (0, 2);
	\vertex (cc2) at (0,-2);
	\vertex  (n1) at (-2, 1.5);
	\vertex  (n2) at (-2, 0.75);
	\vertex  (nb2) at (-2, -1.5);
	\vertex  (nb1) at (-2, -0.75);
	\node at (-2,-1.125) {$\bar{n}$};
	\node at (-2,1.125) {$n$};
	\vertex (g1) at (-1, 1.5);
	\vertex (g2) at (-1, -1.5);
	\vertex (s1) at (-1, .5);
	\vertex (s2) at (-1,-.5);
\diagram* {

(n2)--[ line width = 0.2mm](a2)--[ line width = 0.2mm](e2),
(e1)--[ line width = 0.2mm](b2)--[ line width = 0.2mm](c2)--[ line width = 0.2mm](nb2),
(b1)--[scalar](b2),
(cc1)--[scalar, gray](cc2),
(g1)--[scalar, red, line width = 0.3mm](s1),
(g2)--[scalar, red, line width = 0.3mm](s2),
(s1)--[half right, looseness = 1.75,SGreen, gluon,, line width = 0.3mm](s2) --[half right, looseness = 1.75, SGreen, gluon, line width = 0.3mm](s1),
(n1)--[ line width = 0.2mm](a1)--[ line width = 0.2mm](b1)--[ line width = 0.2mm](c1)--[ line width = 0.2mm](nb1),
};
\end{feynman}
    \filldraw[red] (-1,1.5) ellipse (0.6mm and 1.2mm);
    \filldraw[red] (-1,-1.5) ellipse (0.6mm and 1.2mm);
    \filldraw[red] (-1,.5) ellipse (0.6mm and 1.2mm);
    \filldraw[red] (-1,-.5) ellipse (0.6mm and 1.2mm);
\end{tikzpicture}
}
\end{subfigure}
\begin{subfigure}[b]{0.18\textwidth}
\caption{\qquad \qquad \qquad \qquad }
\centering
\scalebox{.7}{
\begin{tikzpicture}
\begin{feynman}
    \node [crossed dot, NodeBlue] (b1) at (1.5, .5);
    \node [crossed dot, NodeBlue] (b2) at (1.5, -.5);
	\vertex (a1) at (0, 1.5);
	\vertex (c1) at (0,-0.75);
	\vertex (a2) at (0, .75);
	\vertex (c2) at (0,-1.5);
	\vertex (e1) at (0.99,-0.075);
	\vertex (e2) at (0.81, 0.075);
	\vertex (cc1) at (0, 2);
	\vertex (cc2) at (0,-2);
	\vertex  (n1) at (-2, 1.5);
	\vertex  (n2) at (-2, 0.75);
	\vertex  (nb2) at (-2, -1.5);
	\vertex  (nb1) at (-2, -0.75);
	\node at (-2,-1.125) {$\bar{n}$};
	\node at (-2,1.125) {$n$};
	\vertex (g1) at (-1, 1.5);
	\vertex (g2) at (-1, -1.5);
	\vertex (s1) at (-1, 0);
\diagram* {
(s1)--[gluon, SGreen, line width = 0.2mm](b2),
(n2)--[ line width = 0.2mm](a2)--[ line width = 0.2mm](e2),
(e1)--[ line width = 0.2mm](b2)--[ line width = 0.2mm](c2)--[ line width = 0.2mm](nb2),
(b1)--[scalar](b2),
(cc1)--[scalar, gray](cc2),
(g1)--[scalar, red, line width = 0.3mm](g2),
(n1)--[ line width = 0.2mm](a1)--[ line width = 0.2mm](b1)--[ line width = 0.2mm](c1)--[ line width = 0.2mm](nb1),
};
\end{feynman}
    \filldraw[red] (-1,1.5) ellipse (0.6mm and 1.2mm);
    \filldraw[red] (-1,-1.5) ellipse (0.6mm and 1.2mm);
    \filldraw[red] (-1,0) ellipse (0.6mm and 1.2mm););
\end{tikzpicture}
}
\end{subfigure}
\begin{subfigure}[b]{0.18\textwidth}
\caption{\qquad \qquad \qquad \qquad }
\centering
\scalebox{.7}{
\begin{tikzpicture}
\begin{feynman}
    \node [crossed dot, NodeBlue] (b1) at (1.5, .5);
    \node [crossed dot, NodeBlue] (b2) at (1.5, -.5);
	\vertex (a1) at (0, 1.5);
	\vertex (c1) at (0,-0.75);
	\vertex (a2) at (0, .75);
	\vertex (c2) at (0,-1.5);
	\vertex (e1) at (0.99,-0.075);
	\vertex (e2) at (0.81, 0.075);
	\vertex (cc1) at (0, 2);
	\vertex (cc2) at (0,-2);
	\vertex  (n1) at (-2, 1.5);
	\vertex  (n2) at (-2, 0.75);
	\vertex  (nb2) at (-2, -1.5);
	\vertex  (nb1) at (-2, -0.75);
	\node at (-2,-1.125) {$\bar{n}$};
	\node at (-2,1.125) {$n$};
	\vertex (g1) at (-1, 1.5);
	\vertex (g2) at (-1, -.75);
	\vertex (s1) at (-1, 0.375);
\diagram* {
(s1)--[gluon, SGreen, line width = 0.2mm](b2),
(n2)--[ line width = 0.2mm](a2)--[ line width = 0.2mm](e2),
(e1)--[ line width = 0.2mm](b2)--[ line width = 0.2mm](c2)--[ line width = 0.2mm](nb2),
(b1)--[scalar](b2),
(cc1)--[scalar, gray](cc2),
(g1)--[scalar, red, line width = 0.3mm](g2),
(n1)--[ line width = 0.2mm](a1)--[ line width = 0.2mm](b1)--[ line width = 0.2mm](c1)--[ line width = 0.2mm](nb1),
};
\end{feynman}
    \filldraw[red] (-1,1.5) ellipse (0.6mm and 1.2mm);
    \filldraw[red] (-1,-.75) ellipse (0.6mm and 1.2mm);
    \filldraw[red] (-1,.3750) ellipse (0.6mm and 1.2mm););
\end{tikzpicture}
}
\end{subfigure}
\begin{subfigure}[b]{0.23\textwidth}
\caption{\qquad \qquad \qquad \qquad }
\centering
\scalebox{.7}{
\begin{tikzpicture}
\begin{feynman}
    \node [crossed dot, NodeBlue] (b1) at (1.5, .5);
    \node [crossed dot, NodeBlue] (b2) at (1.5, -.5);
	\vertex (a1) at (0, 1.5);
	\vertex (c1) at (0,-0.75);
	\vertex (a2) at (0, .75);
	\vertex (c2) at (0,-1.5);
	\vertex (e1) at (0.99,-0.075);
	\vertex (e2) at (0.81, 0.075);
	\vertex (cc1) at (0, 2);
	\vertex (cc2) at (0,-2);
	\vertex  (n1) at (-2, 1.5);
	\vertex  (n2) at (-2, 0.75);
	\vertex  (nb2) at (-2, -1.5);
	\vertex  (nb1) at (-2, -0.75);
	\node at (-2,-1.125) {$\bar{n}$};
	\node at (-2,1.125) {$n$};
	\vertex (g11) at (-1.5, 1.5);
	\vertex (g12) at (-1.5, -1.5);
	\vertex (g21) at (-.5, 1.5);
	\vertex (g22) at (-.5, -1.5);
\diagram* {
(n1)--[ line width = 0.2mm](a1)--[ line width = 0.2mm](b1)--[ line width = 0.2mm](c1)--[ line width = 0.2mm](nb1),
(n2)--[ line width = 0.2mm](a2)--[ line width = 0.2mm](e2),
(e1)--[ line width = 0.2mm](b2)--[ line width = 0.2mm](c2)--[ line width = 0.2mm](nb2),
(b1)--[scalar](b2),
(cc1)--[scalar, gray](cc2),
(g11)--[scalar, red, line width = 0.3mm](g12),
(g21)--[scalar, red, line width = 0.3mm](g22),
};
\end{feynman}
    \filldraw[red] (-1.5,1.5) ellipse (0.6mm and 1.2mm);
    \filldraw[red] (-1.5,-1.5) ellipse (0.6mm and 1.2mm);
    \filldraw[red] (-.5,1.5) ellipse (0.6mm and 1.2mm);
    \filldraw[red] (-.5,-1.5) ellipse (0.6mm and 1.2mm);
\end{tikzpicture}
}
\end{subfigure}
\begin{subfigure}[b]{0.23\textwidth}
\caption{\qquad \qquad \qquad \qquad }
\centering
\scalebox{.7}{
\begin{tikzpicture}
\begin{feynman}
    \node [crossed dot, NodeBlue] (b1) at (1.5, .5);
    \node [crossed dot, NodeBlue] (b2) at (1.5, -.5);
	\vertex (a1) at (0, 1.5);
	\vertex (c1) at (0,-0.75);
	\vertex (a2) at (0, .75);
	\vertex (c2) at (0,-1.5);
	\vertex (e1) at (0.99,-0.075);
	\vertex (e2) at (0.81, 0.075);
	\vertex (cc1) at (0, 2);
	\vertex (cc2) at (0,-2);
	\vertex  (n1) at (-2, 1.5);
	\vertex  (n2) at (-2, 0.75);
	\vertex  (nb2) at (-2, -1.5);
	\vertex  (nb1) at (-2, -0.75);
	\node at (-2,-1.125) {$\bar{n}$};
	\node at (-2,1.125) {$n$};
	\vertex (g11) at (-1.3, 1.5);
	\vertex (g12) at (-1.3, -1.5);
	\vertex (g21) at (-.7, .75);
	\vertex (g22) at (-.7, -.75);
\diagram* {
(n1)--[ line width = 0.2mm](a1)--[ line width = 0.2mm](b1)--[ line width = 0.2mm](c1)--[ line width = 0.2mm](nb1),
(n2)--[ line width = 0.2mm](a2)--[ line width = 0.2mm](e2),
(e1)--[ line width = 0.2mm](b2)--[ line width = 0.2mm](c2)--[ line width = 0.2mm](nb2),
(b1)--[scalar](b2),
(cc1)--[scalar, gray](cc2),
(g11)--[scalar, red, line width = 0.3mm](g12),
(g21)--[scalar, red, line width = 0.3mm](g22),
};
\end{feynman}
    \filldraw[red] (-1.3,1.5) ellipse (0.6mm and 1.2mm);
    \filldraw[red] (-1.3,-1.5) ellipse (0.6mm and 1.2mm);
    \filldraw[red] (-.7,.75) ellipse (0.6mm and 1.2mm);
    \filldraw[red] (-.7,-.75) ellipse (0.6mm and 1.2mm);
\end{tikzpicture}
}
\end{subfigure}
\begin{subfigure}[b]{0.23\textwidth}
\caption{\qquad \qquad \qquad \qquad }
\centering
\scalebox{.7}{
\begin{tikzpicture}
\begin{feynman}
    \node [crossed dot, NodeBlue] (b1) at (1.5, .5);
    \node [crossed dot, NodeBlue] (b2) at (1.5, -.5);
	\vertex (a1) at (0, 1.5);
	\vertex (c1) at (0,-0.75);
	\vertex (a2) at (0, .75);
	\vertex (c2) at (0,-1.5);
	\vertex (e1) at (0.99,-0.075);
	\vertex (e2) at (0.81, 0.075);
	\vertex (cc1) at (-.75, 2);
	\vertex (cc2) at (-.75,-2);
	\vertex  (n1) at (-2, 1.5);
	\vertex  (n2) at (-2, 0.75);
	\vertex  (nb2) at (-2, -1.5);
	\vertex  (nb1) at (-2, -0.75);
	\node at (-2,-1.125) {$\bar{n}$};
	\node at (-2,1.125) {$n$};
	\vertex (g11) at (-1.5, 1.5);
	\vertex (g12) at (-1.5, -1.5);
	\vertex (g21) at (0, 1.5);
	\vertex (g22) at (0, -1.5);
\diagram* {
(n1)--[ line width = 0.2mm](a1)--[ line width = 0.2mm](b1)--[ line width = 0.2mm](c1)--[ line width = 0.2mm](nb1),
(n2)--[ line width = 0.2mm](a2)--[ line width = 0.2mm](e2),
(e1)--[ line width = 0.2mm](b2)--[ line width = 0.2mm](c2)--[ line width = 0.2mm](nb2),
(b1)--[scalar](b2),
(cc1)--[scalar, gray](cc2),
(g11)--[scalar, red, line width = 0.3mm](g12),
(g21)--[scalar, red, line width = 0.3mm](g22),
};
\end{feynman}
    \filldraw[red] (-1.5,1.5) ellipse (0.6mm and 1.2mm);
    \filldraw[red] (-1.5,-1.5) ellipse (0.6mm and 1.2mm);
    \filldraw[red] (0,1.5) ellipse (0.6mm and 1.2mm);
    \filldraw[red] (0,-1.5) ellipse (0.6mm and 1.2mm);
\end{tikzpicture}
}
\end{subfigure}
\begin{subfigure}[b]{0.23\textwidth}
\caption{\qquad \qquad \qquad \qquad }
\centering
\scalebox{.7}{
\begin{tikzpicture}
\begin{feynman}
    \node [crossed dot, NodeBlue] (b1) at (1.5, .5);
    \node [crossed dot, NodeBlue] (b2) at (1.5, -.5);
	\vertex (a1) at (0, 1.5);
	\vertex (c1) at (0,-0.75);
	\vertex (a2) at (0, .75);
	\vertex (c2) at (0,-1.5);
	\vertex (e1) at (0.99,-0.075);
	\vertex (e2) at (0.81, 0.075);
	\vertex (cc1) at (-.75, 2);
	\vertex (cc2) at (-.75,-2);
	\vertex  (n1) at (-2, 1.5);
	\vertex  (n2) at (-2, 0.75);
	\vertex  (nb2) at (-2, -1.5);
	\vertex  (nb1) at (-2, -0.75);
	\node at (-2,-1.125) {$\bar{n}$};
	\node at (-2,1.125) {$n$};
	\vertex (g11) at (-1.5, 1.5);
	\vertex (g12) at (-1.5, -1.5);
	\vertex (g21) at (0, .75);
	\vertex (g22) at (0, -.75);
\diagram* {
(n1)--[ line width = 0.2mm](a1)--[ line width = 0.2mm](b1)--[ line width = 0.2mm](c1)--[ line width = 0.2mm](nb1),
(n2)--[ line width = 0.2mm](a2)--[ line width = 0.2mm](e2),
(e1)--[ line width = 0.2mm](b2)--[ line width = 0.2mm](c2)--[ line width = 0.2mm](nb2),
(b1)--[scalar](b2),
(cc1)--[scalar, gray](cc2),
(g11)--[scalar, red, line width = 0.3mm](g12),
(g21)--[scalar, red, line width = 0.3mm](g22),
};
\end{feynman}
    \filldraw[red] (-1.5,1.5) ellipse (0.6mm and 1.2mm);
    \filldraw[red] (-1.5,-1.5) ellipse (0.6mm and 1.2mm);
    \filldraw[red] (0,.75) ellipse (0.6mm and 1.2mm);
    \filldraw[red] (0,-.75) ellipse (0.6mm and 1.2mm);
\end{tikzpicture}
}
\end{subfigure}
\caption{2-loop cut diagrams with soft loops.  Diagram (1) is the iterative soft-Glauber graph, diagrams (2) and (3) come from the one-loop amplitude, and diagrams (4) and (5) are the real-emission contributions.  Diagrams (7-10) involve two Glauber exchanges, and cancel when summed over.  Not shown are the graphs given by taking $n\leftrightarrow \bar{n}$.}
\label{fig: 3}
\end{figure}

The graphs which contribute at two loops are shown in figure (2). Graph (1) factors into the product of the one-loop soft-graph and the one loop cut:
\begin{align}
\begin{gathered}
\scalebox{.5}{
\begin{tikzpicture}
\begin{feynman}
    \node [crossed dot, NodeBlue] (b1) at (1.5, .5);
    \node [crossed dot, NodeBlue] (b2) at (1.5, -.5);
	\vertex (a1) at (0, 1.5);
	\vertex (c1) at (0,-0.75);
	\vertex (a2) at (0, .75);
	\vertex (c2) at (0,-1.5);
	\vertex (e1) at (0.99,-0.075);
	\vertex (e2) at (0.81, 0.075);
	\vertex (cc1) at (0, 2);
	\vertex (cc2) at (0,-2);
	\vertex  (n1) at (-2, 1.5);
	\vertex  (n2) at (-2, 0.75);
	\vertex  (nb2) at (-2, -1.5);
	\vertex  (nb1) at (-2, -0.75);
	\node at (-2,-1.125) {$\bar{n}$};
	\node at (-2,1.125) {$n$};
	\vertex (g1) at (-1, .75);
	\vertex (g2) at (-1, -.75);
\diagram* {
(n2)--[ line width = 0.2mm](a2)--[ line width = 0.2mm](e2),
(e1)--[ line width = 0.2mm](b2)--[ line width = 0.2mm](c2)--[ line width = 0.2mm](nb2),
(b1)--[gluon, SGreen, line width = 0.2mm](b2),
(cc1)--[scalar, gray](cc2),
(g1)--[scalar, red, line width = 0.3mm](g2),
(n1)--[ line width = 0.2mm](a1)--[ line width = 0.2mm](b1)--[ line width = 0.2mm](c1)--[ line width = 0.2mm](nb1),
};
\end{feynman}
    \filldraw[red] (-1,.75) ellipse (0.6mm and 1.2mm);
    \filldraw[red] (-1,-.75) ellipse (0.6mm and 1.2mm);
\end{tikzpicture}
}
\end{gathered}
&= - 4 g^4C_F (2C_F -C_A) \int \frac{\dbar^d k_1 \dbar^{d-2}k_{2\perp}\,\left|\frac{2k_1^z}{\nu} \right|^{-\eta}}{k_1^2\,(k_1^+  + i\epsilon)( k_1^- -i\epsilon) \vec{k}_{2\perp}^2} e^{-i\vec{b}_\perp\cdotp(\vec{k}_{1\perp}-\vec{k}_{2\perp})},\nonumber\\
& = \frac{i g^4\, (C_F^2-2 C_FC_A)}{16\pi^2}\frac{\Gamma(1/2-\eta/2)\Gamma(\eta/2)\Gamma(-\varepsilon-\eta/2)\Gamma(-\varepsilon)}{2^{ \eta}\pi^{3/2}\Gamma(1 + \eta/2)}\frac{(\tilde{b}^2\mu^2)^{2\varepsilon + \eta}}{e^{(2\varepsilon + \eta)\gamma_E}},\\
&=i(2C_F^2-C_F  C_A)\frac{\alpha_s^2}{\pi}\bigg[\frac{2\Gamma(-\varepsilon)^2e^{-\varepsilon \gamma_E}}{\eta}(\tilde{b}^2\mu^2)^{\varepsilon} -\frac{1}{\varepsilon^3} +\frac{1}{\varepsilon^2}(L_\nu - L_b) \nonumber\\
&\qquad \qquad \qquad\qquad + \frac{2}{\varepsilon}L_b L_\nu    + \frac23L_b^3 + 2L_b^2 L_\nu  + \zeta(2)(L_b+L_\nu) + \frac{\zeta(3)}{3}\bigg],\nonumber
\end{align}
with $L_\nu = \log\mu^2/\nu^2$. 
The $C_F^2$ term cancels the iteration terms $\gamma_s^{\nu(1)} F^{\dag(1)} -i\pi/2(\gamma_s^{\nu(1)})^2 F^{\dag(0)}$ when added to the cut-renormalization $2\text{Im}[Z_F^{-1}] F^{\dag(0)}$, while the $C_F C_A$ term contributes to the 2-loop rapidity anomalous dimension.  

We next consider the diagrams which contain one-loop corrections to the amplitude, which are the soft ``eye''  and the fermion vacuum polarization graphs.  For the fermion vacuum polarization we have
\begin{align}
\begin{gathered}
\scalebox{.5}{
\begin{tikzpicture}
\begin{feynman}
    \node [crossed dot, NodeBlue] (b1) at (1.5, .5);
    \node [crossed dot, NodeBlue] (b2) at (1.5, -.5);
	\vertex (a1) at (0, 1.5);
	\vertex (c1) at (0,-0.75);
	\vertex (a2) at (0, .75);
	\vertex (c2) at (0,-1.5);
	\vertex (e1) at (0.99,-0.075);
	\vertex (e2) at (0.81, 0.075);
	\vertex (cc1) at (0, 2);
	\vertex (cc2) at (0,-2);
	\vertex  (n1) at (-2, 1.5);
	\vertex  (n2) at (-2, 0.75);
	\vertex  (nb2) at (-2, -1.5);
	\vertex  (nb1) at (-2, -0.75);
	\node at (-2,-1.125) {$\bar{n}$};
	\node at (-2,1.125) {$n$};
	\vertex (g1) at (-1, 1.5);
	\vertex (g2) at (-1, -1.5);
	\vertex (s1) at (-1, .5);
	\vertex (s2) at (-1,-.5);
\diagram* {
(n2)--[ line width = 0.2mm](a2)--[ line width = 0.2mm](e2),
(e1)--[ line width = 0.2mm](b2)--[ line width = 0.2mm](c2)--[ line width = 0.2mm](nb2),
(b1)--[scalar](b2),
(cc1)--[scalar, gray](cc2),
(g1)--[scalar, red, line width = 0.3mm](s1),
(g2)--[scalar, red, line width = 0.3mm](s2),
(s1)--[half right, looseness = 1.75,SGreen, fermion, line width = 0.3mm](s2) --[half right, looseness = 1.75, SGreen, fermion, line width = 0.3mm](s1),
(n1)--[ line width = 0.2mm](a1)--[ line width = 0.2mm](b1)--[ line width = 0.2mm](c1)--[ line width = 0.2mm](nb1),
};
\end{feynman}
    \filldraw[red] (-1,1.5) ellipse (0.6mm and 1.2mm);
    \filldraw[red] (-1,-1.5) ellipse (0.6mm and 1.2mm);
    \filldraw[red] (-1,.5) ellipse (0.6mm and 1.2mm);
    \filldraw[red] (-1,-.5) ellipse (0.6mm and 1.2mm);
\end{tikzpicture}
}
\end{gathered}
& = 2g^2 C_F T_F n_f\int \frac{\dbar^d k_1\dbar^{d-2} k_{2\perp}}{\vec{k}_{2\perp}^2 k_1^2 (k_1 + k_{2\perp})^2}\text{Tr}[\slashed{k_1}\frac{\slashed{\bar{n}}}{2}(\slashed{k}_1 + \slashed{k}_{2\perp})\frac{\slashed{n}}{2}]e^{-i\vec{b}_\perp\cdotp\vec{k}_{2\perp}},\nonumber\\
&= \frac{-i g^4}{4\pi^3} C_F T_F n_f\, \frac{\Gamma(\varepsilon) \Gamma(2-\varepsilon)^2\Gamma(-2\varepsilon)}{\Gamma(4-2\varepsilon)\Gamma(1 + \varepsilon)}\frac{(\tilde{b}^2\mu^2)^{2\varepsilon}}{e^{-2\varepsilon \gamma_E}},\\
&= i \frac{\alpha_s^2}{\pi} C_F T_F n_f\bigg[\frac{1}{3\varepsilon^2} + \frac{1}{\varepsilon}(\frac59 + \frac23L_b) +\frac23L_b^2 + \frac{10}{9}L_b + \frac13\zeta(2) + \frac{28}{27}\bigg].
\end{align}
There is only one other term with the color structure $C_f T_F n_f$, which comes from the two-loop renormalized coupling.  This comes from multiplying the one-loop diagram in eq. (\ref{TMD}) by $Z_{\alpha}$, and gives
\begin{align}
-i C_F\beta_0\frac{\alpha_s^2}{2\pi} \frac{\Gamma(-\varepsilon)}{\varepsilon}(\tilde{b}^2\mu^2)^{\varepsilon}= i\frac{\alpha_s^2}{\pi}C_F\left(\frac{11C_A}{3} - \frac{4 T_F n_f}{3}\right)\bigg[\frac{1}{2\varepsilon^2}+\frac{1}{2\varepsilon}L_b + \frac14L_b^2 + \frac14 \zeta(2)\bigg],
\end{align}
with $\beta_0$ being the 1-loop beta-function coefficient, $\beta_0 = 11/3 C_A - 4/3 T_F n_f$.  Adding the $T_F n_f$-terms, we obtain
\begin{equation}
\text{quark terms}=C_F T_F n_F\frac{\alpha_s^2}{4\pi}\bigg[\frac{-1}{3\varepsilon^2} + \frac{5}{9\varepsilon} -\frac{2}{3}L_b^2 + \frac{10}{9}L_b + \frac{28}{27}\bigg],
\end{equation}
which, after dropping the UV poles and dividing by $-\pi$, is exactly the $T_F n_f$ terms in the two-loop TMD rapidity anomalous dimension.

For the soft-eye graph, we find
\begin{align}
&\begin{gathered}
\scalebox{.5}{
\begin{tikzpicture}
\begin{feynman}
    \node [crossed dot, NodeBlue] (b1) at (1.5, .5);
    \node [crossed dot, NodeBlue] (b2) at (1.5, -.5);
	\vertex (a1) at (0, 1.5);
	\vertex (c1) at (0,-0.75);
	\vertex (a2) at (0, .75);
	\vertex (c2) at (0,-1.5);
	\vertex (e1) at (0.99,-0.075);
	\vertex (e2) at (0.81, 0.075);
	\vertex (cc1) at (0, 2);
	\vertex (cc2) at (0,-2);
	\vertex  (n1) at (-2, 1.5);
	\vertex  (n2) at (-2, 0.75);
	\vertex  (nb2) at (-2, -1.5);
	\vertex  (nb1) at (-2, -0.75);
	\node at (-2,-1.125) {$\bar{n}$};
	\node at (-2,1.125) {$n$};
	\vertex (g1) at (-1, 1.5);
	\vertex (g2) at (-1, -1.5);
	\vertex (s1) at (-1, .5);
	\vertex (s2) at (-1,-.5);
\diagram* {
(n2)--[ line width = 0.2mm](a2)--[ line width = 0.2mm](e2),
(e1)--[ line width = 0.2mm](b2)--[ line width = 0.2mm](c2)--[ line width = 0.2mm](nb2),
(b1)--[scalar](b2),
(cc1)--[scalar, gray](cc2),
(g1)--[scalar, red, line width = 0.3mm](s1),
(g2)--[scalar, red, line width = 0.3mm](s2),
(s1)--[half right, looseness = 1.75,SGreen, gluon,, line width = 0.3mm](s2) --[half right, looseness = 1.75, SGreen, gluon, line width = 0.3mm](s1),
(n1)--[ line width = 0.2mm](a1)--[ line width = 0.2mm](b1)--[ line width = 0.2mm](c1)--[ line width = 0.2mm](nb1),
};
\end{feynman}
    \filldraw[red] (-1,1.5) ellipse (0.6mm and 1.2mm);
    \filldraw[red] (-1,-1.5) ellipse (0.6mm and 1.2mm);
    \filldraw[red] (-1,.5) ellipse (0.6mm and 1.2mm);
    \filldraw[red] (-1,-.5) ellipse (0.6mm and 1.2mm);
\end{tikzpicture}
}
\end{gathered}
\begin{aligned}
 &= 2g^4 C_F C_A\int \frac{\dbar^{d-2}k_{2\perp}\dbar^d k_1\,\left|\frac{2k_1^z}{\nu} \right|^{-\eta}}{\vec{k}_{2\perp}^2(k_1 + k_{2\perp}) k_1^2}\bigg[\frac{4 [k_1\cdotp(k_{2\perp} + k_1)]^2}{n\cdotp k_1\,\bar{n}\cdotp k_1} \nonumber\\
&\qquad  + \left\{(d-2)n\cdotp k_1\,\bar{n}\cdotp k_1 + 4k_{2\perp}^2 -2(k_1 + k_{2\perp})^2 -2 k_1^2\right\}\bigg] e^{-i\vec{b}_\perp\cdotp\vec{k}_{2\perp}},\end{aligned}\nonumber\\
&= -i\frac{g^4C_F C_A}{8\pi^3}\bigg[\frac{\Gamma(1/2-\eta/2)\Gamma(-\varepsilon-\eta/2)\Gamma(-2\varepsilon-\eta/2)}{4^{ -\varepsilon}\eta\,\Gamma(1/2-\varepsilon-\eta/2)}\frac{(\tilde{b}^2\mu^2)^{2\varepsilon + \eta/2}}{e^{(2\varepsilon +\eta/2) \gamma_E}}\left(\frac{\mu^2}{\nu^2}\right)^{-\eta/2} 
\nonumber\\
&\qquad \qquad  + \frac{\Gamma(\varepsilon)\Gamma(-2\varepsilon)}{2\Gamma(1 +\varepsilon)}\left(\frac{\Gamma(2-\varepsilon)^2}{\Gamma(4-2\varepsilon)}-2\frac{\Gamma(1-\varepsilon)^2}{\Gamma(2-2\varepsilon)}\right)\frac{(\tilde{b}^2\mu^2)^{2\varepsilon + \eta/2}}{e^{-(2\varepsilon +\eta/2) \gamma_E}}\bigg],\label{SoftEyeTMD}\\
&= -i\frac{C_F C_A\alpha_s^2}{4\pi}\bigg[\bigg\{\frac{4\Gamma(-\varepsilon)^2e^{-2\varepsilon\gamma_E}}{\eta} + \frac{3}{\varepsilon^3} + \frac{2}{\varepsilon^2}\left(2L_b -L_\nu\right)+\frac{1}{\varepsilon}\left(2L_b^2-4L_b L_\nu + 3\zeta(2)\right)\nonumber\\
&\qquad \qquad \qquad  -4L_b^2 L_\nu  + 4\zeta(2) L_b-2\zeta(2)L_\nu + 6\zeta(3) \bigg\}-\bigg\{\frac{11}{3\varepsilon^2}+\frac{1}{3\varepsilon}\left(\frac{67}{3} +22 L_b\right)\nonumber\\ 
&\qquad\qquad \qquad  +\frac{22}{3}L_b^2 + \frac{134}{9}L_b + \frac{11}{3}\zeta(2)+ \frac{404}{27}\bigg\}\bigg].\nonumber
\end{align}
On the final line we have written the $\eta$- and $\varepsilon$-expansions such that the terms in the first set of curly brackets come from the rapidity divergent term the line above, and the terms in the second set of square brackets all come from the second set of curly brackets.

In principle, we should also take into account the flower graph, which is given by contracting the two soft gluon emission vertex off a Glauber with itself (see \cite{Rothstein2016AnViolation} for the appropriate Feynman rule), however this diagram is scaleless and thus 
can be ignored .
\begin{equation}
\begin{gathered}
    \scalebox{.5}{
\begin{tikzpicture}
\begin{feynman}
    \node [crossed dot, NodeBlue] (b1) at (1.5, .5);
    \node [crossed dot, NodeBlue] (b2) at (1.5, -.5);
	\vertex (a1) at (0, 1.5);
	\vertex (c1) at (0,-0.75);
	\vertex (a2) at (0, .75);
	\vertex (c2) at (0,-1.5);
	\vertex (e1) at (0.99,-0.075);
	\vertex (e2) at (0.81, 0.075);
	\vertex (cc1) at (0, 2);
	\vertex (cc2) at (0,-2);
	\vertex  (n1) at (-2, 1.5);
	\vertex  (n2) at (-2, 0.75);
	\vertex  (nb2) at (-2, -1.5);
	\vertex  (nb1) at (-2, -0.75);
	\node at (-2,-1.125) {$\bar{n}$};
	\node at (-2,1.125) {$n$};
	\vertex (g1) at (-1.3333, 1.5);
	\vertex (g2) at (-1.3333, -1.5);
	\vertex (s1) at (-1, .5);
	\vertex (s2) at (-1,-.5);
	\vertex (s) at (-1.3333,0);
\diagram* {
(n2)--[ line width = 0.2mm](a2)--[ line width = 0.2mm](e2),
(e1)--[ line width = 0.2mm](b2)--[ line width = 0.2mm](c2)--[ line width = 0.2mm](nb2),
(b1)--[scalar](b2),
(cc1)--[scalar, gray](cc2),
(g1)--[scalar, red, line width = 0.3mm](g2),
(n1)--[ line width = 0.2mm](a1)--[ line width = 0.2mm](b1)--[ line width = 0.2mm](c1)--[ line width = 0.2mm](nb1),
};
    \draw (s) [gluon, SGreen, line width = 0.3mm]arc [start angle=-180, end angle=180, radius=0.5cm];
\end{feynman}
    \filldraw[red] (-1.3333,1.5) ellipse (0.6mm and 1.2mm);
    \filldraw[red] (-1.3333,-1.5) ellipse (0.6mm and 1.2mm);
    \filldraw[red] (-1.3333,0) ellipse (0.6mm and 1.2mm);
\end{tikzpicture}
}
\end{gathered}= 4g^4 C_F C_A\int \frac{\dbar^{d-2}k_{2\perp}\dbar^d k_1\,\left|\frac{2k_1^z}{\nu} \right|^{-\eta}}{k_1^2 \,\vec{k}_{2\perp}^2n\cdotp k_1\,\bar{n}\cdotp k_2}e^{-i \vec{b}_\perp\cdotp \vec{k}_{2\perp}}.
\end{equation}

There are two different real emission graphs which involve the Lipatov vertex (the coupling of a soft gluon to the Glauber).  The first of these is given by
\begin{align}
 &\begin{gathered}
\scalebox{.5}{
\begin{tikzpicture}
\begin{feynman}
    \node [crossed dot, NodeBlue] (b1) at (1.5, .5);
    \node [crossed dot, NodeBlue] (b2) at (1.5, -.5);
	\vertex (a1) at (0, 1.5);
	\vertex (c1) at (0,-0.75);
	\vertex (a2) at (0, .75);
	\vertex (c2) at (0,-1.5);
	\vertex (e1) at (0.99,-0.075);
	\vertex (e2) at (0.81, 0.075);
	\vertex (cc1) at (0, 2);
	\vertex (cc2) at (0,-2);
	\vertex  (n1) at (-2, 1.5);
	\vertex  (n2) at (-2, 0.75);
	\vertex  (nb2) at (-2, -1.5);
	\vertex  (nb1) at (-2, -0.75);
	\node at (-2,-1.125) {$\bar{n}$};
	\node at (-2,1.125) {$n$};
	\vertex (g1) at (-1, 1.5);
	\vertex (g2) at (-1, -1.5);
	\vertex (s1) at (-1, 0);
\diagram* {
(s1)--[gluon, SGreen, line width = 0.2mm](b2),
(n2)--[ line width = 0.2mm](a2)--[ line width = 0.2mm](e2),
(e1)--[ line width = 0.2mm](b2)--[ line width = 0.2mm](c2)--[ line width = 0.2mm](nb2),
(b1)--[scalar](b2),
(cc1)--[scalar, gray](cc2),
(g1)--[scalar, red, line width = 0.3mm](g2),
(n1)--[ line width = 0.2mm](a1)--[ line width = 0.2mm](b1)--[ line width = 0.2mm](c1)--[ line width = 0.2mm](nb1),
};
\end{feynman}
    \filldraw[red] (-1,1.5) ellipse (0.6mm and 1.2mm);
    \filldraw[red] (-1,-1.5) ellipse (0.6mm and 1.2mm);
    \filldraw[red] (-1,0) ellipse (0.6mm and 1.2mm););
\end{tikzpicture}
}
\end{gathered}
\begin{aligned}&= -2g^4 C_F C_A\int \frac{\dbar \bar{n}\cdotp k_1\dbar n\cdotp k_2 \dbar^{d-2}k_{1\perp}\dbar^{d-2}k_{2\perp}\left|\frac{\bar{n}\cdotp k_1 + n\cdotp k_2}{\nu} \right|^{-\eta}}{\vec{k}_{1\perp}^2\vec{k}_{2\perp}^2\, \bar{n}\cdotp k_1 n\cdotp k_2}e^{-i \vec{b}_\perp\cdotp \vec{k}_{1\perp}}\nonumber\\
&\qquad \qquad \times\left(-2\bar{n}\cdotp k_1 n\cdotp k_2 -2\vec{k}_{1\perp}^2 -2\vec{k}_{2\perp}^2\right) \delta^{(+)}(-\bar{n}\cdotp k_1 n\cdotp k_2 -(\vec{k}_{1\perp} -\vec{k}_{2\perp})^2),
\end{aligned}\nonumber\\
&=i \frac{C_F C_A g^4}{4\pi^3}\frac{\Gamma(\eta/2)\Gamma(1-\varepsilon)\Gamma(-\varepsilon-\eta/2)}{(\eta + 4\varepsilon)\Gamma(1 +\eta)}\frac{(\tilde{b}^2\mu^2)^{2\varepsilon + \eta/2}}{e^{(2\varepsilon +\eta/2) \gamma_E}}\left(\frac{\mu^2}{\nu^2}\right)^{-\eta/2},\\
&= -i C_F C_A \frac{\alpha_s^2}{2\pi}\bigg[\frac{4\Gamma(-\varepsilon)^2e^{-2\varepsilon\gamma_E}}{\eta} + \frac{3}{\varepsilon^3} + \frac{2}{\varepsilon^2}(2L_b-L_\nu)  + \frac{1}{\varepsilon}(2L_b^2 -4 L_b L_\nu + \zeta(2))\nonumber\\
&\qquad \qquad \qquad\qquad  -4L_b^2 L_\nu-2\zeta(2)L_\nu\bigg].
\end{align}
The second single real emission vertex is given by
\begin{align}
&\begin{gathered}
\scalebox{.5}{
\begin{tikzpicture}
\begin{feynman}
    \node [crossed dot, NodeBlue] (b1) at (1.5, .5);
    \node [crossed dot, NodeBlue] (b2) at (1.5, -.5);
	\vertex (a1) at (0, 1.5);
	\vertex (c1) at (0,-0.75);
	\vertex (a2) at (0, .75);
	\vertex (c2) at (0,-1.5);
	\vertex (e1) at (0.99,-0.075);
	\vertex (e2) at (0.81, 0.075);
	\vertex (cc1) at (0, 2);
	\vertex (cc2) at (0,-2);
	\vertex  (n1) at (-2, 1.5);
	\vertex  (n2) at (-2, 0.75);
	\vertex  (nb2) at (-2, -1.5);
	\vertex  (nb1) at (-2, -0.75);
	\node at (-2,-1.125) {$\bar{n}$};
	\node at (-2,1.125) {$n$};
	\vertex (g1) at (-1, 1.5);
	\vertex (g2) at (-1, -.75);
	\vertex (s1) at (-1, 0.375);
\diagram* {
(s1)--[gluon, SGreen, line width = 0.2mm](b2),
(n2)--[ line width = 0.2mm](a2)--[ line width = 0.2mm](e2),
(e1)--[ line width = 0.2mm](b2)--[ line width = 0.2mm](c2)--[ line width = 0.2mm](nb2),
(b1)--[scalar](b2),
(cc1)--[scalar, gray](cc2),
(g1)--[scalar, red, line width = 0.3mm](g2),
(n1)--[ line width = 0.2mm](a1)--[ line width = 0.2mm](b1)--[ line width = 0.2mm](c1)--[ line width = 0.2mm](nb1),
};
\end{feynman}
    \filldraw[red] (-1,1.5) ellipse (0.6mm and 1.2mm);
    \filldraw[red] (-1,-.75) ellipse (0.6mm and 1.2mm);
    \filldraw[red] (-1,.3750) ellipse (0.6mm and 1.2mm););
\end{tikzpicture}
}
\end{gathered}
\begin{aligned}
&=- g^4 C_F C_A\int \frac{\dbar \bar{n}\cdotp k_1\dbar n\cdotp k_2 \dbar^{d-2}k_{1\perp}\dbar^{d-2}k_{2\perp}\left|\frac{\bar{n}\cdotp k_1 + n\cdotp k_2}{\nu} \right|^{-\eta}}{\vec{k}_{1\perp}^2\vec{k}_{2\perp}^2\, \bar{n}\cdotp k_1 n\cdotp k_2}e^{-i \vec{b}_\perp\cdotp( \vec{k}_{1\perp} -\vec{k}_{2\perp})}\nonumber\\
&\qquad \qquad \times \left(-2\bar{n}\cdotp k_1 n\cdotp k_2 -2\vec{k}_{1\perp}^2 -2\vec{k}_{2\perp}^2\right)\delta^{(+)}(-\bar{n}\cdotp k_1 n\cdotp k_2 -(\vec{k}_{1\perp} -\vec{k}_{2\perp})^2),\end{aligned}\nonumber\\
&=-2i g^4 C_F C_A \frac{\Gamma(\eta/2)\Gamma(-\varepsilon)^2\Gamma(1 + \varepsilon)\Gamma(-2\varepsilon-\eta/2)}{2^{6 + \eta}\pi^{5/2}\Gamma(1/2 + \eta/2)\Gamma(-2\varepsilon)\Gamma(1+\varepsilon+\eta/2)}\frac{(\tilde{b}^2\mu^2)^{2\varepsilon + \eta/2}}{e^{(2\varepsilon +\eta/2) \gamma_E}}\left(\frac{\mu^2}{\nu^2}\right)^{-\eta/2},\\
&=i C_F C_A\frac{\alpha_s^2}{4\pi}\bigg[\frac{4\Gamma(-\varepsilon)^2e^{-2\varepsilon\gamma_E}}{\eta} -\frac{1}{\varepsilon^3} +\frac{2}{\varepsilon^2} L_\nu+\frac{1}{\varepsilon}\left(2 L_b^2 + 4 L_b L_\nu + \zeta(2)\right)\nonumber\\
&\qquad \qquad \qquad \qquad  +\frac{8}{3}L_b^3+4 L_b^2 L_\nu+4 \zeta(2) L_b +2\zeta(2)L_\nu +\frac{28}{3}\zeta(3) \bigg]. 
\end{align}

Adding up all the (non-abelian) terms gives the final result
\begin{align}
&C_F C_A + C_F T_F n_f \,\text{terms}\nonumber\\
&=i C_F C_A \frac{\alpha_s^2}{4\pi}\bigg[\frac{11}{3\varepsilon^2} -\frac{67}{9\varepsilon} + \frac{2\zeta(2)}{\varepsilon} -\frac{11}{3}L_b^2 + 4\zeta(2)L_b -\frac{134}{9}L_b + 14\zeta(3) -\frac{404}{27}\bigg] \nonumber\\
&\qquad  +i C_F T_F n_f \frac{\alpha_s^2}{4\pi}\bigg[-\frac{4}{3\varepsilon^2} + \frac{20}{9\varepsilon} +\frac43 L_b^2 + \frac{40}{9}L_b + \frac{112}{27}\bigg].
\end{align}
We see that all the $1/\eta$ poles, $\log \nu$ terms, $1/\varepsilon^3$ UV divergences, and non-local divergence terms have all cancelled in the sum over diagrams, as expected.  Dropping the $1/\varepsilon$ poles gives exactly $ -i\pi$ times the two-loop TMD rapidity anomalous dimension:
\begin{eqnarray}
\gamma_\nu^s = -C_F C_A \frac{\alpha_s^2}{4\pi^2}\bigg[  -\frac{11}{3}L_b^2 + 4\zeta(2)L_b -\frac{134}{9}L_b + 14\zeta(3) -\frac{404}{27}\bigg]   - C_F T_F n_f \frac{\alpha_s^2}{4\pi^2}\bigg[\frac43 L_b^2 + \frac{40}{9}L_b + \frac{112}{27}\bigg]  \nonumber \\
\end{eqnarray}
This agrees with the known result \cite{Echevarria2016UniversalNNLO}.


Lastly, we note that the rapidity anomalous dimension for the TMD soft function has been calculated to four loops in \cite{PhysRevLett.129.162001, Moult2022TheDimension}.  This calculation was accomplished by using the correspondence between soft and rapidity anomalous dimensions \cite{Vladimirov2016Soft-/rapidity-Correspondence}.  It would certainly be interesting to explore any implications this has in the context of the work presented here.

\section{Conclusions}

In this paper we have shown that all the large logs that show up in a certain class of S matrix element are controlled by the phase. 
This includes RG logs of invariant mass ratios, as was first shown \cite{Caron-Huot2016RenormalizationS-matrix}, as well as large logs
of rapidity ratios. We have demonstrated how one can calculate the rapidity anomalous dimensions
for both local and non-local operators at two loops. By focusing on the S-matrix phase we are able to extract these anomalous dimensions by calculating the much simpler set of cut diagrams. Furthermore, since we are calculating the rapidity finite anomalous dimensions directly, instead of having to calculate counter-terms, the integrals do not need a rapidity regulator, which, in general, makes integrating more challenging. If one chooses to calculate using Feynman diagrams, as opposed to using on-shell methods, then individual diagrams will in general need to be regulated, whereas finite integrals will only arise once one combines diagrams. In the case of the massive Sudakov form factor there was no need to
a rapidity regulator even at the diagramatic level.
While there as no new results in this paper, we believe that the formalism developed here will be useful for other cases where the
higher order corrections have yet to be calculated such as the Regge trajectory.  However, as we will show in a forthcoming paper 
the applications of the ideas presented here will need to be changed slightly when calculating forward scattering.

\vskip1in 
\noindent {\bf Acknowledgments:} 
The authors benefitted from discussions with Ian Moult, Duff Neill,  Iain Stewart and HuaXing Zhu. We would also like to thank
Julio Parra-Martinez for making us aware of Ref. \cite{Caron-Huot2016RenormalizationS-matrix}, and Simon Caron-Huot for making us aware of \cite{Fadin:1995xg}.
This work is supported by the US Department of Energy (HEP) Award DE-SC0013528. IZR would like to thank the  Erwin-Schrödinger-Institute for Mathematics and Physics
for hospitality.

\appendix
\section{Canonical Calculation of RAD}
\subsubsection{The Massive Gluon Form Factor}
The one loop calculation of the MGFF RAD was carried out in detail in \cite{Chiu2012ATheory} and will be recapitulated here in order to help the reader appreciate how this present methodology simplifies matters. 
As usual in SCET we break up the calculation into the three relevant regions of momentum space, hard, soft, and collinear. For the processes we will be interested in the IR (soft and collinear) modes have the same invariant mass
which is of order $M^2$. 
The factorized form factor is written as \cite{Bauer:2002nz}
\begin{equation}
\mathcal{F} = \left(\bar{\xi}_n W_n\right) S_n^\dag\Gamma S_{\bar{n}} (W_{\bar{n}}^\dag \xi_{\bar{n}}),
\end{equation}
$W_n$ and $S_n$ are collinear and soft Wilson lines respectively.

At tree level one has
\begin{equation}
F(p_1, p_2) = \bra{p_1, p_2} \mathcal{F} \ket{0} = \bar{u}_n(p_2)\Gamma v_{\bar{n}}(p_1)\equiv S_\Gamma.
\end{equation}
The one loop corrections get contributions from the soft and collinear ($n$ and $\bar n$) regions.
As discussed above, the Glauber region is captured by the soft and can be safely ignored in this context.
The soft contribution is given by 
\beq
\label{softs}
I_S=g^2 C_F S_\Gamma \left[-\frac{e^{\gamma_E \epsilon}\Gamma(\epsilon) \left(\frac{\mu}{M}\right)^{2 \epsilon}}{4\pi^2\eta}
+\frac{1}{4\pi^2}\left( \frac{\ln (\frac{\mu}{\nu})}{\epsilon} 
+\ln ^2(\frac{\mu}{M}) -2\ln (\frac{\mu}{M})\ln (\frac{\nu}{M})+ \frac{1}{2\epsilon^2}\right)-\frac{1}{96} \right]
\eeq
while the collinear contribution is given by
\beq
\label{cols}
I_n=g^2 C_F S_\Gamma \left[\frac{e^{\gamma_E \epsilon}\Gamma(\epsilon) \left(\frac{\mu}{M}\right)^{2 \epsilon}}{8\pi^2\eta}+\frac{1}{4\pi^2}\left( \ln (\frac{\mu}{M}) \ln (\frac{\nu}{\bar n \cdot p_1}) +\ln (\frac{\mu}{M}) +\frac{1}{2\epsilon} \left(1+\ln (\frac{\nu }{\bar n \cdot p_1}) \right)+\frac{1}{2}\right)-\frac{1}{48}\right] \, ,
\eeq
and $I_{\bar n}$ by replacing $\bar n \cdot p_1$ with $n\cdot p_2$.

Here $\eta$ is a rapidity regulator which is necessary as the individual contributions are ill defined without it, despite the use of dimensional regularization and the gluon mass. As previously mentioned, this regulator can be introduced in a gauge invariant way as described above equation (\ref{reg}), where $\nu$ is the rapidity factorization scale which separates the soft and collinear modes. This regulator breaks boost invariance, which is regained when all three sectors
are added together.

Summing the sectors we find 
\beq
I_S+I_{\bar n}+I_n=g^2 S_\Gamma C_F \left[\frac{1}{4\pi^2}\left(\frac{1}{2\epsilon^2}+ \frac{\ln (\frac{\mu}{Q})}{\epsilon}+\frac{1}{\epsilon}+\ln ^2(\frac{\mu}{M})+2\ln (\frac{\mu}{M})+2 \ln \frac{M}{\mu} \ln \frac{Q}{M}+1 \right)-\frac{5}{96}\right] \, ,
\eeq
where we have used $\bar n \cdot p_1 = n\cdot p_2 = Q$. 
We see that in the sum the $\eta$ (rapidity) divergences vanish as promised and there is no dependence on the scale $\nu$ and the answer is boost invariant. The rapidity anomalous dimensions for the collinear ($n$) and soft ($S$) sectors can then be extracted by
\beq
\label{def}
\gamma_\nu^{n,S} = -Z^{-1}_{n,S} \frac{\partial}{\partial \ln\nu} Z_{n,S} \, ,
\eeq
where $Z_{n,S}$ are the sector renormalization constants
\begin{align}
Z_S&=1-\frac{g(\mu)^2 w^2 C_F}{4\pi^2} \left[\frac{e^{\epsilon \gamma_E}\Gamma(\epsilon) \left(\frac{\mu}{M}\right)^{2 \epsilon}}{\eta}-\frac{1}{2\epsilon^2}-\frac{\ln \frac{\mu}{\nu}}{\epsilon}\right] \, , \nonumber \\
Z_{n}&=1+\frac{g(\mu)^2 w^2 C_F}{4\pi^2} \left[\frac{e^{\epsilon\gamma_E }\Gamma(\epsilon) \left(\frac{\mu}{M}\right)^{2 \epsilon}}{2\eta}+\frac{1}{2\epsilon}\bigg(1+\ln \frac{\nu}{\nbar \cdot p_1} \bigg)\right] \, ,
\end{align}
$Z_{\bar n}$ follows by replacing $n\leftrightarrow \bar n$ and $p_1 \leftrightarrow p_2$.
$Z_\psi$ is wave function renormalization which is the same as in full QCD.
\beq
Z_{\psi}=1-\frac{g(\mu)^2C_F}{16\pi^2 \epsilon}.
\eeq

Using the fact that the book-keeping parameter $w$ (which gets set to one at the end of the calculation) obeys \footnote{This is analogue of
$\mu \frac{d}{d\mu}(\mu^{-2 \epsilon}g)= -2 \epsilon \mu^{-2 \epsilon}g+...$. Only here the coupling does not depend upon $\nu$}
\beq
\nu\frac{dw}{d\nu }=-\frac{\eta}{2} w.
\eeq

The RAD are then given by
\bea
\label{A1}
\gamma_\nu^S &=& \frac{\alpha}{\pi}C_F \ln(M^2/\mu^2) \nn \\
\gamma_\nu^n &=& -\frac{\alpha}{2\pi}C_F \ln(M^2/\mu^2),
\eea
and obey the consistency relation $\gamma_\nu^S+2\gamma_\nu^n=0$ which ensures that the
form factor is independent of $\nu$. Notice that the Glauber played no role in this calculation. 
The reason for this is that the soft-Glauber zero bin exactly cancels the Glauber contribution (see section (10) of \cite{Rothstein2016AnViolation}).
The Glauber region is what generates the imaginary part.

\subsubsection{The Massive Quark Form Factor}

The MQFF is not an IR finite quantity, unlike the massive gluon case. The full theory UV finite result is given by (\cite{Bernreuther:2004ih})\footnote{This reference gives the full two loop result.}
\beq
\label{full}
F^{(1)}_{YM}= \frac{\alpha_s C_F}{4 \pi}(\frac{2 \ln (\frac{-Q^2}{m^2}-i \epsilon)-2}{\epsilon_{\text{IR}}}
-(2 \ln(\frac{m^2}{\mu^2})-3)
- \ln^2 (\frac{-Q^2}{m^2}-i \epsilon)
+2 \ln(\frac{m^2}{\mu^2}) -4 +\frac{\pi^2}{3})
\eeq
The $i\pi$ in the first log should be reproduced in the EFT while the imaginary part of the double log is part of the
hard matching coefficient \cite{Hoang2015HardLoops}.
Nonetheless we can use the RAD to resum IR divergences which is of formal interest. Here we will calculate the one loop MQFF RAD
using the canonical method directly from the soft rapidity anomalous dimensions following from direct calculation.

We see in the first term an IR divergence that must be reproduced in the EFT. Moreover this
divergence is proportional to $\ln(Q)$. How this Log is reproduced in the EFT is not obvious since
the EFT loops are naively independent of $Q$. The collinear loop ($ \bar n \cdot k \sim \bar n \cdot p^\prime \sim n \cdot p \sim Q$) is given by
\beq
I_c(m)=-2i g^2 C_F S_\Gamma \int \frac{[d^dk]}{(k^2+i \epsilon)} \frac{ \bar n \cdot (p^\prime-k)}{(k^2- n \cdot p^\prime \bar n \cdot k-\bar n \cdot p^\prime n \cdot k+i\epsilon)} \frac{ n \cdot p}{(- n \cdot p \bar n \cdot k+i\epsilon)}
\eeq
with the other collinear mode, with $k || p^\prime$, giving the same result with $n \leftrightarrow \bar n$.
One might think that the sum of the collinear modes could reproduce a $\ln(Q)$. However, an explicit calculation
shows that this is not the case, at least naively. Performing the $n\cdot k$ integrals by contours leaves
\beq
I_c(m)= \frac{g^2}{4\pi} C_F\frac{\Omega_{2-2 \epsilon}}{(2\pi)^{2-2\epsilon}}\int_0^1 d\xi \int [dk_\perp]\frac{k_\perp^{1-2 \epsilon}}{(k_\perp^2+m^2\xi^2)}\frac{(1-\xi)}{\xi}
\eeq
This integral needs no rapidity regulator, which seems to be problematic since the soft, scaleless, integral
is rapidity divergent. Expanding, the result is given by
\beq
I_c(m)=\frac{\alpha C_F}{4 \pi}\left( 
\frac{ 2}{ \epsilon ^2}+\frac{ 4}{ \epsilon }-2\frac{ \ln(\frac{m^2}{\mu^2})}{ \epsilon }
+8+\frac{\pi^2 }{6}+ \ln^2(\frac{m^2}{\mu^2}) -4 \ln(\frac{m^2}{\mu^2})\right),
\eeq
which does not reproduce the proper $Q$ dependence mentioned above.
This issue is resolved only once we properly separate out the IR and UV divergences in this
integral.
To do so 
we split the collinear integral as follows
\beq
I_c(m)=(I_c(m)-I_c(0))+I_c(0).
\eeq
The term in the parentheses is a UV finite, but IR divergent integral, whereas the last term is rapidity divergent and has both IR and UV divergences. 
The result for the term in the parenthesis is given by
\bea
I_c(m)-I_c(0)&=&-\frac{g^2}{4\pi} C_F\frac{\Omega_{2-2 \epsilon_{\text{IR}}}}{(2\pi)^{2-2\epsilon_{\text{IR}}}}\int_0^1 d\xi \int [dk_\perp]\frac{k_\perp^{1-2 \epsilon_{\text{IR}}}}{k_\perp^2(k_\perp^2+m^2\xi^2)}(m^2\xi)(1-\xi)
\nn \\
&=&\frac{\alpha C_F}{4 \pi}\left( 
\frac{ 2}{ \epsilon_{\text{IR}} ^2}+\frac{ 4}{ \epsilon_{\text{IR}} }-2\frac{ \ln(\frac{m^2}{\mu^2})}{ \epsilon_{\text{IR}} }
+8+\frac{\pi^2 }{6}+ \ln^2(\frac{m^2}{\mu^2}) -4 \ln(\frac{m^2}{\mu^2})\right) \eea
as expected we just reproduce $I_c(m)$ but now with a well-defined notion of $\epsilon$. 

Splitting the result into UV and IR pieces, $I_c(0)$ is given by
\bea
I_c(0)&=&-\frac{g^2 C_F}{(2\pi)}\frac{\Omega_{2-2 \epsilon}}{(2\pi)^{2-2\epsilon}}(\frac{\bar n \cdot p^\prime}{\nu})^{-\eta} \int_0^1 d \xi \int dk_\perp \frac{k_\perp^{1-2\epsilon}}{k_\perp^2} (1-\xi)\xi^{-1-\eta} \nn \\
&=& -\frac{g^2 C_F}{(2\pi)}\frac{\Omega_{2-2 \epsilon}}{(2\pi)^{2-2\epsilon}}(\frac{\bar n \cdot p^\prime}{\nu})^{-\eta} \int_0^1 d \xi \int dk_\perp k_\perp^{1-2\epsilon} (1-\xi)\xi^{-1-\eta}(\frac{1}{k_\perp^2+M^2}+\frac{M^2}{k_\perp^2(k_\perp^2+M^2)}).\nn \\
\eea
In the first/second terms we replace $\epsilon \rightarrow \epsilon_{\text{IR}}/\epsilon_{\text{UV}}$.
Upon adding the collinear diagram from the other sector and subtracting the $1/\eta$ pole we are left with
\beq
I_c(0)=-\frac{\alpha C_F}{ 2\pi} \frac{1}{ \epsilon_{\text{UV}}}
(\ln \left(\frac{Q^2}{\nu^2 }\right)-2)+\frac{\alpha C_F}{ 2\pi} \frac{1}{ \epsilon_{\text{IR}}}
(\ln \left(\frac{Q^2}{\nu^2 }\right)-2),
\eeq
which reproduces the troubling IR divergent $\ln(Q)$ term in (\ref{full}). Though it seems to be missing the
imaginary contribution, which is part of the soft integral, or equally, from the Glauber once the Glauber zero bin has
been subtracted from the soft.

The soft integral vanishes since it is scaleless. 
However, that cannot be the end of the story since we are still missing the imaginary part of the IR divergence in (\ref{full}).
We have to split the soft piece into its UV and IR divergent pieces.
The soft integral is given by
\beq
I_s=2 g^2 C_F \Gamma \int \frac{[d^{d}k_\perp][dk_z]\mid \frac{k_z}{\nu} \mid^{-\eta}}{k^2+i\epsilon} \frac{1}{n \cdot k +i \epsilon} \frac{1}{\bar n \cdot k -i \epsilon},
\eeq
performing the energy integral by contours leaves
\beq
I_s=2 i g^2 C_F \Gamma \int [d^{d-2}k_\perp][dk_z] \mid \frac{k_z}{\nu} \mid^{-\eta}
\left[
\frac{1}{((k_z^2+k_\perp^2)^{1/2}+i\epsilon)}\frac{1}{k_\perp^2 +i \epsilon} 
- 
\frac{1}{(k_\perp^2+i\epsilon)}\frac{1}{k_z-i \epsilon} 
\right]
\eeq
The first term is manifestly real while the second will generate the proper phase to complete (\ref{full}).
However to see this we must split the second term into UV and IR pieces after which one generates a factor
of $i \pi(\frac{1}{\epsilon_{\text{UV}}}-\frac{1}{\epsilon_{\text{IR}}})$. The IR term when added to the collinear term generates
the minus sign in the argument of the log in (\ref{full}), while the UV term will cancel the UV divergence in the
hard piece ensuring finiteness of the full theory result. The imaginary part of the log squared term in (\ref{full}) comes
from the hard contribution.

Then applying eq.(\ref{def}) to the rapidity pole pieces of $I_s$ gives
\beq
\label{canon}
\gamma^S_\nu=\frac{\alpha_s C_F}{\pi}\frac{1}{\epsilon_{\text{IR}}}.
\eeq
Notice that all of the UV divergences have cancelled as required. This anomalous dimension will sum
the terms of the form $\frac{1}{\epsilon_{\text{IR}}} \ln(Q^2)$. The terms of the form $\ln^2(Q^2)$ will be summed using
the usual RG, via the cusp anomalous dimension.



\bibliography{references.bib}

\end{document}